\documentclass[
  reprint,
  amssymb,
  aps,
  pra,
  floatfix,
  superscriptaddress
]{revtex4-2}

\usepackage{amsmath}
\usepackage{booktabs}
\usepackage{comment}
\usepackage{glossaries}
\usepackage{graphicx}
\usepackage[version=4]{mhchem}
\usepackage{siunitx}
\usepackage[monochrome]{color}
\usepackage{hyperref}
\usepackage{tabularx}

\hypersetup{colorlinks, 
	linkcolor={blue!75!black!80!yellow},
	citecolor={blue!75!black!80!yellow}, 
	urlcolor={blue!75!black!80!yellow}
}

\usepackage[normalem]{ulem}
\newcommand{\Cadd}[1]{#1}

\newacronym{dft}{DFT}{density functional theory}
\newacronym{md}{MD}{molecular dynamics}
\newacronym{ml}{ML}{machine learning}
\newacronym{nep}{NEP}{neuroevolution potential}
\newacronym{pes}{PES}{potential energy surface}
\newacronym{qedft}{QEDFT}{quantum electrodynamical density-functional theory}

\begin{document}

\title{Machine Learning for Polaritonic Chemistry: Accessing chemical kinetics}

 \author{Christian Sch\"afer}
 \email[Electronic address:\;]{christian.schaefer.physics@gmail.com}
 \affiliation{
 Department of Physics, Chalmers University of Technology, 412 96 G\"oteborg, Sweden}
 \affiliation{
  Department of Microtechnology and Nanoscience, MC2, Chalmers University of Technology, 412 96 G\"oteborg, Sweden}
 \author{Jakub Fojt}
 \affiliation{
 Department of Physics, Chalmers University of Technology, 412 96 G\"oteborg, Sweden}

 \author{Eric Lindgren}
 \affiliation{
 Department of Physics, Chalmers University of Technology, 412 96 G\"oteborg, Sweden}

  \author{Paul Erhart}
   \email[Electronic address:\;]{erhart@chalmers.se}
  \affiliation{
  Department of Physics, Chalmers University of Technology, 412 96 G\"oteborg, Sweden}

\date{\today}
\begin{abstract}
Altering chemical reactivity and material structure in confined optical environments is on the rise, and yet, a conclusive understanding of the microscopic mechanisms remains elusive. 
This originates mostly from the fact that accurately predicting vibrational and reactive dynamics for soluted ensembles of realistic molecules is no small endeavor, and adding (collective) strong light-matter interaction does not simplify matters.
Here, we establish a framework based on a combination of machine learning (ML) models, trained using density-functional theory calculations, and molecular dynamics to accelerate such simulations.
We then apply this approach to evaluate strong coupling, changes in reaction rate constant, and their influence on enthalpy and entropy for the deprotection reaction of 1-phenyl-2-trimethylsilylacetylene, which has been studied previously both experimentally and using \textit{ab initio} simulations.
While we find qualitative agreement with critical experimental observations, especially with regard to the changes in kinetics, we also find differences in comparison with previous theoretical predictions.
The features for which the ML-accelerated and \textit{ab initio} simulations agree show the experimentally estimated kinetic behavior.
Conflicting features indicate that a contribution of \Cadd{dynamic} electronic polarization to the reaction process is more relevant then currently believed.
Our work demonstrates the practical use of ML for polaritonic chemistry, discusses limitations of common approximations and paves the way for a more holistic description of polaritonic chemistry.
\end{abstract}

\maketitle


\section{Introduction}

If confined electromagnetic fields interact sufficiently strongly with matter, their excitations hybridize and give rise to new quasi-particles called polaritons \cite{genet2021inducing, garcia2021manipulating, simpkins2021mode, sidler2021perspective, doi:10.1021/acs.chemrev.2c00788, mandal2023}.
Strong light-matter coupling has been used to alter chemical reactivity \cite{hutchison2012, Munkhbat2018, ahn_herrera_simpkins_2022, chen2022cavity, galego2016, fregoni2020strong, schafer2021shining}, for which the term polaritonic chemistry has been coined.
Vibrational strong coupling in particular is a promising candidate for practical application, demonstrating the inhibition \cite{thomas2016, thomas2020ground, ahn_herrera_simpkins_2022, doi:10.1021/acsphotonics.3c00243}, steering \cite{Thomas2019}, and catalysis \cite{singh2023solvent,wiesehan2021negligible} of chemical processes at room temperature.
Especially appealing features include the ability to non-intrusively control the path of the chemical reaction by adjusting external parameters, such as the distance between mirrors, and its existence in the absence of any externally provided energy.
The latter sets polaritonics apart from Floquet engineering which typically suffers from heating and uncontrolled dissipation processes~\cite{schafer2018insights}.
Besides the specific control of chemical reactivity, polaritonics has been shown to give 
rise to a myriad of effects that range from commanding single molecules \cite{chikkaraddy2016, wang2017coherent, Ojambati2019, flick2017, haugland2020coupled, arul2022giant}, over altering energy transfer \cite{coles2014b, orgiu2015, zhong2016, fukushima2022inherent, wellnitz2022disorder, herrera2016,schafer2019modification, groenhof2019tracking, csehi2022competition, campos2019resonant, li2021collective, kumar2023extraordinary, stumper2023localization, doi:10.1021/acs.nanolett.3c00867, timmer2023plasmon, cui2023comparing, engelhardt2023polariton}, to the control of phase transitions in extended systems \cite{peter2005exciton, liu2015strong, jarc2023cavity, latini2021ferroelectric, lenk2022dynamical}.

Delivering a conclusive theoretical understanding for vibrational strong coupling has remained difficult.
Especially the experimentally observed resonance dependence in combination with an increase of rate changes for increasing emitter concentration, and clear trends in chemical kinetics are critical features that a theoretical model should capture.
Initial attempts based on standard transition-state theory \cite{galego2019cavity, campos2020polaritonic, litao2020} failed to reproduce any significant frequency dependence.
Active development along the lines of Grote-Hynes \cite{li2020resonance, sun2022suppression,sun2023modification} and Pollak–Grabert–H\"anggi \cite{lindoy2022} theory showed some frequency dependence but a connection to experiments has remained unsuccessful.
A recent work by Sch\"afer \textit{et al.}~\cite{schafer2021shining} tackled the problem from first principles by utilizing \gls{qedft} \cite{ruggenthaler2014, tokatly2013, flick2018cavity, schafer2021making} in combination with a self-consistent update of the nuclear motion according to Ehrenfest's equation for the experimentally investigated deprotection reaction of 1-phenyl-2-trimethylsilylacetylene (PTA)~\cite{thomas2016, thomas2020ground}.
This \textit{ab initio} theory recovered critical components of the frequency dependence and suggested a microscopic theory based on adjusted vibrational energy redistribution for the cavity induced modification of chemical reactivity.
Chen \textit{et al.} found experimental support for this hypothesis using 2D spectroscopy \cite{chen2022cavity}.
Nonetheless, a prediction of kinetic quantities remained inaccessible given the computational cost of \gls{qedft} in combination with Ehrenfest dynamics.

In this work, we establish a framework that combines \gls{ml} models, trained on data from \gls{dft} calculations, with \gls{md} simulations to arrive at a more efficient, yet accurate description of the experimentally relevant S$_N$2 reaction of PTA \cite{thomas2016} under strong coupling to a cavity.
We find a pronounced frequency dependence for the chemical reaction rate constant along with changes in reaction enthalpy and entropy that are qualitatively consistent with experiment.
Interestingly, we discover frequency domains outside the experimentally validated window for which the present \gls{ml}-accelerated approach predicts a rate constant enhancing character in contrast to earlier fully \textit{ab initio} simulations~\cite{schafer2021shining}, which rather suggest inhibition.
Here, we tentatively attribute this difference to the simplifications inherent to the present \gls{md} approach, suggesting that the latter, despite being widely used, has relevant limitations in its applicability to polaritonic chemistry.
Further investigations, beyond the scope of the present work, will be needed to provide a more detailed understanding, which we expect to also generate useful insight into the microscopic mechanisms. 

\section{Methodology}

Nonrelativistic quantum electrodynamics commonly starts at the minimal coupling Hamiltonian in the Coulomb gauge \cite{loudon1988,craig1998, schafer2020relevance, doi:10.1021/acs.chemrev.2c00788}, where all charged particles couple via longitudinal Coulomb interaction to each other and to the transverse vector potential ($\nabla \cdot \textbf{A} = 0$) 
\begin{align*}
    \hat{H} &= \sum_i^{N_e+N_n} \frac{1}{2m_i} \big( -i\hbar \nabla_i - q_i \hat{\textbf{A}}(\textbf{r}_i) /c \big)^2 \\
    &+ \frac{1}{8\pi\varepsilon_0} \sum_{i,j}^{N_e+N_n} \frac{q_i q_j}{\vert \textbf{r}_i - \textbf{r}_j\vert}\\
    &+ \frac{\varepsilon_0}{2}\int d\textbf{r} \big[ \hat{\textbf{E}}_\perp(\textbf{r})^2 + c^2 \hat{\textbf{B}}(\textbf{r})^2 \big].  \notag
\end{align*}
Polaritonic chemistry comes in different flavors that are distinguished by the frequency of the confined modes and the physical nature of the resonator, e.g., plasmonic, Fabry-P\'erot, or whispering gallery cavities, which influences their fundamental coupling strength and quality factor.
The accurate description of a realistic optical environment is a challenge in itself~\cite{buhmann2013dispersionI,lentrodt2020ab,feist2020macroscopic,schafer2021shortcut,schaeferembedding2022}, but simple single-mode models often suffice to obtain a qualitative understanding of the relevant emitter dynamics.

The goal of this work is to provide a qualitative investigation of vibrational strong coupling and its influence on chemical reactivity for experimentally relevant molecules.
In particular, we focus on kinetic changes, its impact on enthalpy and entropy, and the consequences of simplifications in the molecular dynamics.
For this reason, we stay conceptually close to the previous work by Sch\"afer \textit{et al.} \cite{schafer2021shining} and rely on the Born-Oppenheimer approximation, projecting on the electronic ground-state and ignoring electronic polarizations induced by the cavity, such that the effective nuclear-photonic Hamiltonian takes the form \cite{loudon1988, flick2018cavity, schafer2020relevance, li2020cavity}
\begin{align*}
    \hat{H} &= \hat{T}_\text{nuclei} + \hat{V}_\text{PES} + \hbar\omega_c ( \hat{a}^\dagger \hat{a} + \tfrac{1}{2}) \\
    &- \sqrt{\tfrac{\hbar \omega_c}{2\varepsilon_0 V_c}} ( \boldsymbol\varepsilon_c \cdot \hat{\boldsymbol{\mu}})(\hat{a}^\dagger + \hat{a}) + \tfrac{1}{2\varepsilon_0 V_c}( \boldsymbol\varepsilon_c \cdot \hat{\boldsymbol{\mu}})^2.
\end{align*}
The nuclei couple self-consistently to a single electromagnetic mode of the cavity, treated classically as the mode displacement $q_c(t) = \sqrt{\frac{\hbar}{2\omega_c}}\langle \hat{a}^\dagger + \hat{a}\rangle$ in the following, with frequency $\omega_c$, effective cavity volume $V_c$, fixed polarization $\boldsymbol\varepsilon_c$, and molecular dipole moment $ \hat{\boldsymbol{\mu}}$.
Taking the classical limit for the nuclei, the system follows standard Hamiltonian mechanics with forces originating from the Poisson bracket $\{\textbf{p}_j,\mathcal{H}\}$.
In order to see appreciable changes for the dynamics of a single molecule, we choose large coupling values $g_0 = e a_0 \sqrt{\omega_c/2\hbar\varepsilon_0 V_c}$ that are consistent with Ref.~\citenum{schafer2021shining}.
We refer the interested reader to Refs.~\cite{schafer2021shining, sidler2020polaritonic, schaeferembedding2022, sidler2023unraveling,schnappinger2023cavity} for an extended discussion on the potential motivation of effective single-molecule coupling values.
The chosen coupling is considerably larger than experimentally achievable values in Fabry-P\'erot cavities but the qualitative agreement with experiments suggests similar microscopic mechanisms.
Our results and discussion can be partially transferred to plasmonic cavities which feature substantial single-molecule couplings \cite{chikkaraddy2016,Ojambati2019}.

The \emph{electronic} force acting on nucleus $j$ are obtained from the \gls{pes} according to
\begin{align}
    \label{eq:pes-force}
    \textbf{F}^{j}_{\text{PES}} = -\nabla_j V_{\text{PES}}(\textbf{r}),
\end{align}
and contributes to the total force $\textbf{F}^j=\textbf{F}^j_{\text{PES}} + \textbf{F}^j_{c}$ together with the \emph{optical} force, which can be computed from the derivative of the dipole moment vector
\begin{align}
    \label{eq:cavity-force}
    \textbf{F}^j_{c} =  \frac{1}{\sqrt{\varepsilon_0 V_c}} \nabla_j \big[\boldsymbol\varepsilon_c \cdot \boldsymbol{\mu}\big] \bigg( \omega_c q_c(t) - \frac{1}{\sqrt{\varepsilon_0 V_c}} \boldsymbol\varepsilon_c \cdot \boldsymbol{\mu} \bigg).
\end{align}
The only interaction between light and matter arises then via the time evolution of the photonic mode $q_c(t)$ and the gradient of the projected dipole moment.
The cavity mode displacement $q_c(t)$ depends on the history of the dipole moment (see SI \Cadd{Sec.~II A} for details)
\begin{align*}
    q_c(t) =
         q_c(0) \cos(\omega_c t)
         + \int_0^t \frac{\varepsilon_c\cdot\boldsymbol{\mu}(t')}{\sqrt{\varepsilon_0 V_c}}  \sin(\omega_c (t - t'))\:\mathrm{d}t',
\end{align*}
where vanishing of the initial momentum $\dot{q}_c(0)=0$ is explicitly enforced.

We leverage the modular character of the forces by implementing a custom (``cavity'') propagator using the \textsc{ase} Python package \cite{larsen2017atomic}.
The calculator requires merely the initial molecular configuration as well as estimators for the forces arising from the \gls{pes} (Eq.~\eqref{eq:pes-force}) and the dipole moment vector (Eq.~\eqref{eq:cavity-force}).
Possible estimators may include machine-learned models, empirical force fields, and even \textit{ab initio} calculations based on, e.g., \gls{dft}.
We note that this approach can also be easily combined with the embedding radiation-reaction approach \cite{schaeferembedding2022} in the future, thus providing an elegant path for including collective coupling and realistic optical environments.

\begin{figure*}[ht]
    \centering
    \includegraphics[width=1.0\linewidth]{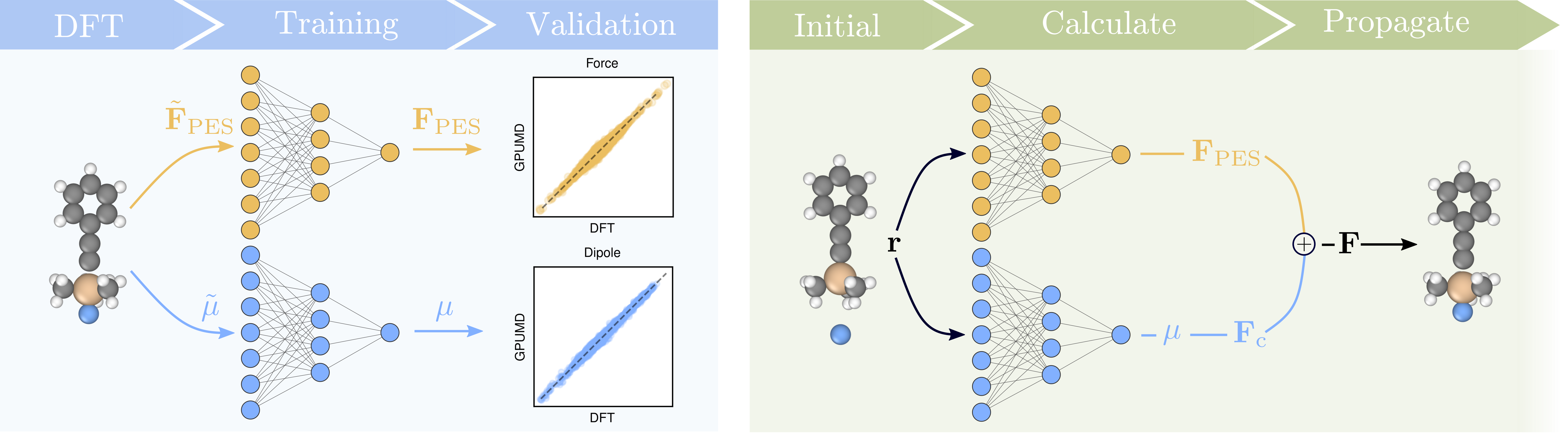}
    \caption{
        \textbf{Methodological flow-chart.}
        \Acrfull{dft} is used to calculates energies, forces, and dipoles.
        Positional information of a structure is translated into descriptor space and \acrlong{nep} models for the \acrfull{pes} and the dipole moment vector $\boldsymbol{\mu}$ are trained using supervised learning (left-hand side).
        The final models are combined to compute the effective forces acting on the nuclei that are then used to propagate the system in time (right-hand side).
    }
    \label{fig:concept}
\end{figure*}

Obtaining forces and dipole derivatives thus represents the main obstacle in the \gls{md} approach, as in reality a molecule can easily pass through tens of thousands of configurations before a reaction occurs.
One possible, but practically often too expensive, approach is to perform \textit{ab initio} electronic structure calculations at each point, usually referred to as \textit{ab intio} \gls{md} \cite{marx2009ab}.
While the cost of a single \gls{dft} calculation might be comparably low, the sheer quantity of calculations required for a statistically meaningful evaluation, i.e., obtaining thousands of trajectories with tens of thousands of \gls{dft} evaluations each, is highly prohibitive and practically limits simulation time as well as system and ensemble size.
Another prominent approach is based on empirical force fields \cite{van1990force, ponder2003force}, which are computationally orders of magnitude more efficient than \gls{dft} calculations.
They are, however, restricted with respect to accuracy as well as availability, and offer limited transferability.
Even if we would have a suitable force field for a system, there is no guarantee that, e.g., a force field fitted in an aqueous solution will perform well for simulations in vacuum and vice versa.

As we will show in the following, employing \gls{ml} techniques in combination with \gls{md} provides a feasible, predictive, and scalable path, striking a balance between computational cost and accuracy.

\subsection{Electronic and optical forces using neural networks}

In the present work, we developed two different models using the \gls{nep} framework as implemented in the \textsc{gpumd} package \cite{fan2021, fan2022, fan2022gpumd}\Cadd{. O}ne for predicting the \gls{pes} $V_\text{PES}(\textbf{r})$\Cadd{, as well as the associated force $\textbf{F}_\text{PES}(\textbf{r})$,} and one for predicting dipole moments $\boldsymbol{\mu}(\textbf{r})$  (\autoref{fig:concept}; left).
We refer the reader to the SI for an extensive discussion of the training and testing procedure of both models. 

The \gls{nep} models are then combined to obtain the total force used to propagate the system in time with a custom integrator implemented in the \textsc{ase} package \cite{larsen2017atomic} (\autoref{fig:concept}; right). 
We refer the interested reader to Refs.~\cite{fan2022gpumd, xu2023dipole} for a more extensive presentation of the \gls{nep} framework and its application to tensorial quantities.

The \gls{nep} approach employs a simple forward bias multi-layer perceptron with a single hidden layer in combination with a flexible descriptor to predict the atomic energy $U_i$ for each atom in a system, by decomposing the total energy into individual contributions from each atom, $U = \sum_i U_i$.
The model consists of a fully connected network with a single hidden layer, yielding the following expression for the predicted energy,
\begin{equation}
    U_i = \sum_{\mu=1}^{N_\text{neu}} w_\mu^{(1)} \textrm{tanh} \left( \sum_{v=1}^{N_{des}} w_{\mu v}^{(0)} q_v^i - b_\mu^{(0)}  \right) - b^{(1)}.    
\end{equation}
The two weight matrices  $w_{\mu v}^{(0)}$ and $w_\mu^{(1)}$ are the weights for the input and hidden layer, with $b_\mu^{(0)}$ and $b^{(1)}$ being their respective biases, and $\textrm{tanh}$ is used as the activation function for the input layer.
The so-called descriptor vector $q_v^i$ of length $N_\text{des}$ indexed by $v$ can be seen as representation of the local chemical environment of atom $i$, is a function of the pairwise distances $\mathbf{r}_{ij} = \mathbf{r}_i - \mathbf{r}_j$, and serves as input to the network. 
If $q_v^i$ uniquely describes a molecular configuration is determined by a basis expansion over N-body interactions within a cutoff radius $r_c$. The expansion is truncated at 4-body interactions, which is sufficient to accurately describe local changes.
A key feature of the \gls{nep} formalism is that these descriptors also contain trainable parameters, which allows the network to tailor the descriptors more individually to different atomic configurations. 
Predictions for forces and virials can be obtained by computing the gradient of the predicted site energies, i.e., the \gls{pes} force acting on atom $i$ is
\begin{equation}
    \mathbf{F}_{i} = \sum_{i\neq j} \frac{\partial U_i}{\partial \mathbf{r}_{ij}} - \frac{\partial U_j}{\partial \mathbf{r}_{ji}}.
\end{equation}
Additionally, the NEP formalism may be extended to predict other tensorial properties such as dipole moments, which are obtained as 
\begin{align}
    \boldsymbol{\mu} &= - \sum_i^N \sum_{j\neq i} \mathbf{r}_{ij} \cdot 
    \left(\mathbf{r}_{ij} \otimes \frac{\partial U_i}{\partial \mathbf{r}_{ij}}\right) \nonumber \\
    &= - \sum_i \sum_{j\neq i} r_{ij}^2 
    \left(\frac{\partial U_i}{\partial \mathbf{r}_{ij}}\right).
    \label{eq:dipole-prediction}
\end{align}

Having replaced the estimates for forces and dipoles with our \gls{nep} models, we are now equipped to address the question how an optical resonator might influence the here discussed S$_N$2 reaction.

\section{Results and discussion}

In a first step, we perform ensemble calculations with preserved particle number, volume, and temperature (NVT) in absence of the cavity to obtain a reference value for the transition-state enthalpy of $\Delta H^\ddagger = \qty{0.345}{\electronvolt}=\qty{33.3}{\kilo\joule\per\mol}$.
The latter is in excellent agreement with experimental estimates of $\Delta H^\ddagger = 35 \pm 4$~kJmol$^{-1}$ \cite{thomas2020ground}.
\Gls{dft} calculations using the nudged elastic band approach \Cadd{in combination with a transition-state optimization} provide a higher barrier of \qty{0.43}{\electronvolt} 
\Cadd{. This illustrates}
the limitation of estimating the enthalpy from the \qty{0}{\kelvin} energy difference between minimum and transition state \Cadd{and the significance of vibrational contributions}. 
Detailed information as well as various benchmarks can be found in the SI.
We define a reaction event when the relevant Si--C bond is stretched beyond \qty{3.5}{\angstrom}, which exceeds the transition-state Si--C distance of approximately \qty{3}{\angstrom}, to avoid counting eventual recrossing events.
If not further specified, all observables are obtained from ensemble averages involving \num{1000} trajectories.
They are initialized with Boltzmann sampled velocities (kept fixed when changing cavity parameters) at the nonequilibrium \ce{F^-}~+~PTA state used in Ref.~\citenum{schafer2021shining}, and propagated preserving particle number, volume, and energy (NVE).
The cavity displacement is selected such that no electric field exists at time zero, i.e., the cavity force is zero.
The initial state features an energy difference to the minimum \ce{PTAF^-} configuration of \qty{1.34}{\electronvolt}.
This shortens the necessary calculation time and avoids spurious interplay between thermostat and cavity.
One should note, however, that it also limits the transferability of the obtained rate constants to experimental observations.
Nonetheless, we can extract changes in rate constant and thus contribute valuable insight to the current hypothesis behind polaritonic chemistry.


\subsection{Strong Coupling}

\begin{figure*}[t]
    \includegraphics[width=1\textwidth]{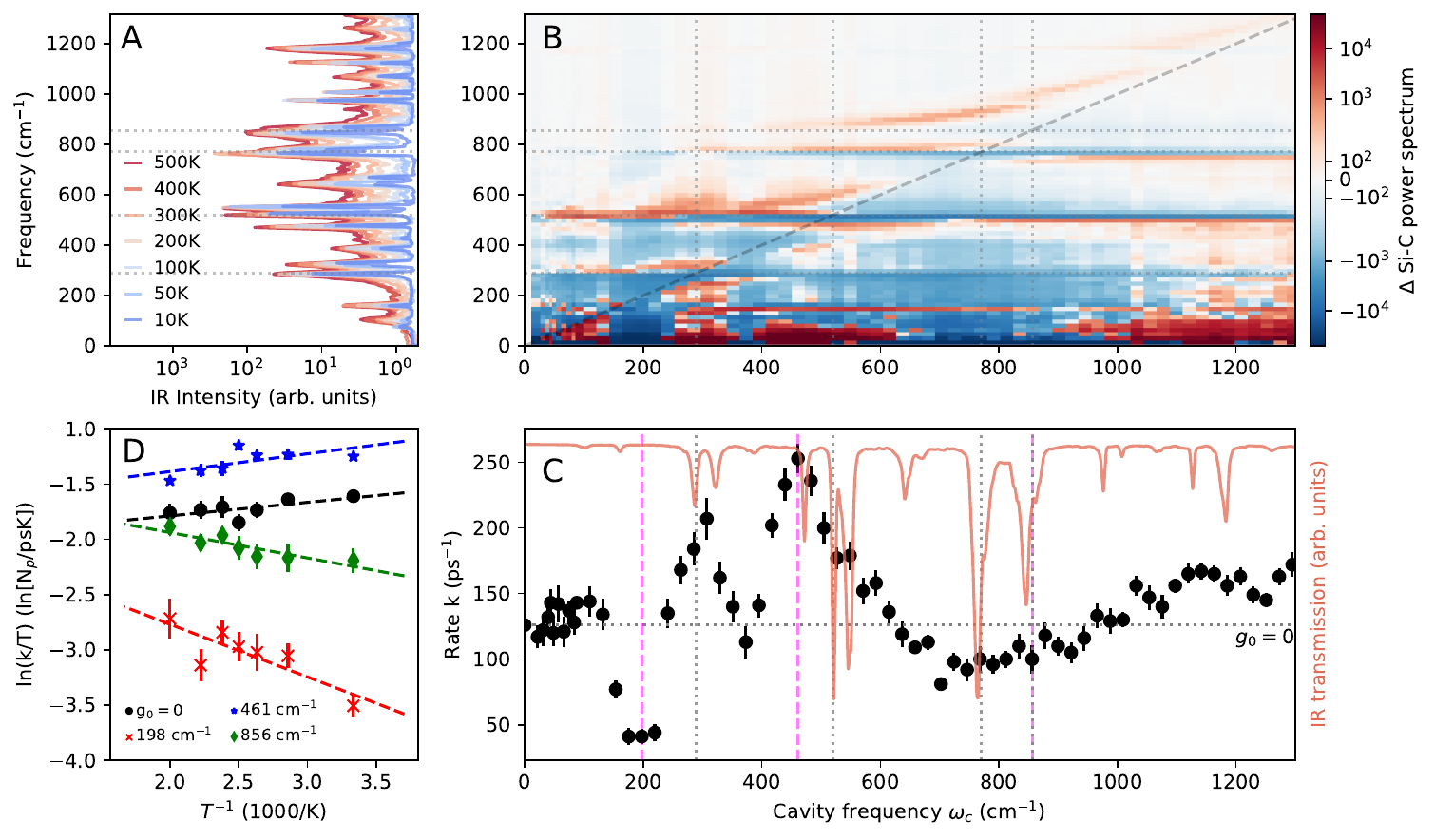}%
    \caption{
    \textbf{Modification of reaction rate constants by coupling to the cavity.}
    (A) Temperature-dependent infrared spectrum outside the cavity, projected on the cavity polarization.
    (B) Difference in power spectra of the Si--C bond for PTAF$^-$ coupled to the cavity with $g_0/\omega_c=1.132$ at 400~K vs the same system outside the cavity.
    Gray dotted lines ($\hbar\omega$ = 290, 520, 770, 856~cm$^{-1}$) serve as guides to the eye.
    (C) Rate constant (black dots) and standard error (black bars) for the unidirectional reaction $\text{PTAF}^- \rightarrow \text{FtMeSi}+\text{PA}^{-}$ at 400~K and $g_0/\omega_c=1.132$.
    The rate constant is calculated as number of products after 2~ps.
    Transmission spectrum at 400~K (red line, defined as negative of absorption spectrum).
    Vertical magenta-colored dashed lines indicate relevant frequencies used for the kinetic estimates in (D).
    (D) Eyring plot in free space and for relevant cavity resonances given in inverse cm with $g_0/\omega_c=1.132$.
    The extracted change in enthalpy and entropy is collected in \autoref{tab:thermodata}. Negative enthalpy originates from the selected initial state and should only be interpreted relative to the equilibrium enthalpy (see text).
    The rate constant is calculated as number of products $N_p$ per ps.
    }
    \label{fig:IRpowerSiC}
\end{figure*}

Strong coupling requires the existence of optically active vibrational modes near the cavity frequency.
Certainly, the vibrational spectrum is sensitive to temperature, as illustrated in \autoref{fig:IRpowerSiC}A.
The reaction-defining Si--C bond contributes fractionally to most vibrational excitations but primarily at frequencies below 1300~cm$^{-1}$.
\autoref{fig:IRpowerSiC}B illustrates the corresponding power spectrum during the reaction process at \qty{400}{\kelvin} with strong coupling to the cavity.
We keep the ratio $g_0/\omega_c$ constant in our calculations.
Changing the cavity length $L_c$ ($V_c = A_c L_c$), which is the experimental way to tune the frequency of the Fabry-P\'erot cavity, leads in a simplified modes picture to $g_0 \propto \sqrt{\omega_c}/\sqrt{L_c}$ and $\omega_c \propto 1/L_c$, i.e., a larger distance between the mirrors reduces both frequency and coupling strength with $1/L_c$.
Following the gray-dashed diagonal $\omega = \omega_c$, we can clearly identify multiple avoided crossings (hybridizations) with substantial Si--C contribution.
Each of the avoided crossings contributes with additional low-energy components (following the vertical gray dotted lines), suggesting comparably slow changes, i.e., on the timescale of the reaction.
The reorganization of the methyl groups and the proper F--Si--C alignment (bending modes) are critical steps in the reaction pathway.
Their interplay with the cavity manifests in the constant contribution at $\hbar\omega\approx \qty{117}{\per\centi\meter}$, which is further detailed in \autoref{sec:limitations}.

\subsection{Resonance dependence}

Intuitively, we expect the low-frequency Si--C contribution to play a major role for the reactivity as the defining Si--C bond breaking step requires a few hundred femto seconds.
Given the fact that the contribution changes non-monotonically, it is not surprising that the rate constant also changes non-monotonically (\autoref{fig:IRpowerSiC}C).
We observe pronounced regions of inhibited reactivity, especially around \qty{200}{\per\centi\meter} and the domain including the 770 and \qty{856}{\per\centi\meter} vibrations.
Furthermore, a near twofold enhancement of the rate constant at around 290 and \qty{460}{\per\centi\meter} is visible.
Experimental investigations are only available near 856 cm$^{-1}$ and support the general inhibiting trend in this domain, albeit featuring an inhibiting effect only in a narrow frequency window.
A non-monotonous rate change, however, is in conflict with previous \textit{ab initio} simulations in Ref.~\citenum{schafer2021shining}.
Interestingly, the most pronounced feature at 200 cm$^{-1}$ seems not related to any optical excitation and is also not connected to the curvature at the transition state, which we estimate with both our \gls{nep} model and \gls{dft} calculations to be approximately 73 cm$^{-1}$.

We will elaborate the conceptual differences to Ref.~\citenum{schafer2021shining} and possible explanations in \autoref{sec:limitations}, but it should be noted that a strict comparison is difficult due to the difference in observable, temperature, and a substantial difference in statistical sampling.
The slow but steady increase in rate constant for large frequencies could partially originate from numerical deviations in the finite-differences approximation of the dipole gradient (see conservation of energy in SI).
With this in mind, let us disentangle the mechanism behind the catalysing and inhibiting effects by estimating enthalpic and entropic changes.

\subsection{Kinetics}
\label{sec:kinetics}

We repeat calculations for the rate constant for various temperatures at four frequencies extracted from \autoref{fig:IRpowerSiC}C which are characteristic for their respective frequency domains.
The results are collected in an Eyring plot (\autoref{fig:IRpowerSiC}D) and indicate conceptually different mechanisms for inhibition and rate constant enhancement.
The corresponding Arrhenius plot is provided in the SI.
We would like to emphasize that those changes originate from the independent dynamics of an ensemble of trajectories and the Eyring equation
\begin{align*}
    k = \kappa \frac{k_B T}{h}e^{\tfrac{\Delta S^\ddagger}{k_B}}e^{-\tfrac{\Delta H^\ddagger}{k_B T}}
\end{align*}
is used only to enable a comparison with the experimentally extracted enthalpy and entropy.
Changes in the transmission coefficient $\kappa$ thus contribute to an altered entropy.

Without cavity environment ($\omega_c=0$, black), we estimate a weak but negative enthalpic barrier $\Delta H$.
This should be seen as the change induced by the chosen initial state and NVE conditions, i.e., the elevated initial state will provide almost no energetic barrier and yet, the cavity will alter this ``barrier''. We demonstrate in SI \Cadd{Sec.~II B} that performing rate constant estimates under NVT conditions and sufficient equilibration time provides accurate rates in agreement with the experiment outside the cavity, i.e., our methodology provides the correct kinetics outside the cavity but the chosen initial state shifts the enthalpy up in energy.
We collect the changes in enthalpy $\Delta\Delta H$ and entropy $\Delta\Delta S $ in relation to the cavity-free NVE results in \autoref{tab:thermodata}.

\begin{table}[b]
    \centering
    \caption{
        Change in enthalpy and entropy compared to free-space Eyring result (black data in \autoref{fig:IRpowerSiC}D).
    }
    \label{tab:thermodata}
    \begin{tabular}{lrr}
    \toprule
    $ \omega_c$ (cm$^{-1}$)
    & \multicolumn{1}{c}{$\Delta\Delta H$ (eV)}
    & \multicolumn{1}{c}{$\Delta\Delta S $ ($k_B$)} \\
    \midrule
    198  &  $+0.052$  &  +0.22 \\ 
    461  &  $-0.003$  &  +0.32 \\ 
    856  &  $+0.030$  &  +0.56 \\ 
    \bottomrule
    \end{tabular}
\end{table}

At the inhibiting frequencies 198 cm$^{-1}$ (red) and 856 cm$^{-1}$ (green), the enthalpy $\Delta H$ increases considerably.
The $\omega_c = 198$ cm$^{-1}$ excitation, without optically active vibrational mode support, shows a weak increase in entropy, suggesting that the dynamic effect of the cavity is to ``simply'' raise the transition-state barrier.
Even though $\omega_c =856$ cm$^{-1}$ (green) shows an overall smaller change, entropic changes are large compared to the other frequencies, which suggests that the cavity assigns a slightly stronger dissociative character to the reaction.
Both features are in qualitative agreement with experiment \cite{thomas2016, thomas2020ground}.
Taking into account that the correct enthalpy obtained from our NVT calculations in free space is $\Delta H^\ddagger = 0.345$~eV, the inhibiting frequencies render the reaction more temperature sensitive, which can lead to enhanced rates for large temperatures.
Such a trend in temperature sensitivity has been widely observed in experiments~\cite{thomas2016, thomas2020ground, singh2023solvent}.

At the catalysing frequency $\omega_c = 461$ cm$^{-1}$ on the other hand there is almost no change in enthalpy, but a noticeable change in entropy, suggesting that the mechanism dominating here is not related to the experiments which typically showed a clear change in enthalpy.
It is important to emphasize, that utilization of the Eyring equation is especially problematic in this domain as a further increase in reaction speed implies that the trajectory will spend less time around the reactant well.
This in turn implies that the kinetic arguments underlying the Eyring equation, i.e., a separation of time scales between reactant equilibration and transmission process, become questionable since the transition-state is then part of the equilibration process.
Nonetheless, a similar offset in rate constant without change in enthalpy has been observed when employing Grote-Hynes rate-theory \cite{doi:10.1021/acs.jpclett.1c01847}.
All three domains relate to different kinetic changes and thus suggest slightly different mechanisms.

\begin{figure*}[t]
    \centering
    \includegraphics[width=1.0\textwidth]{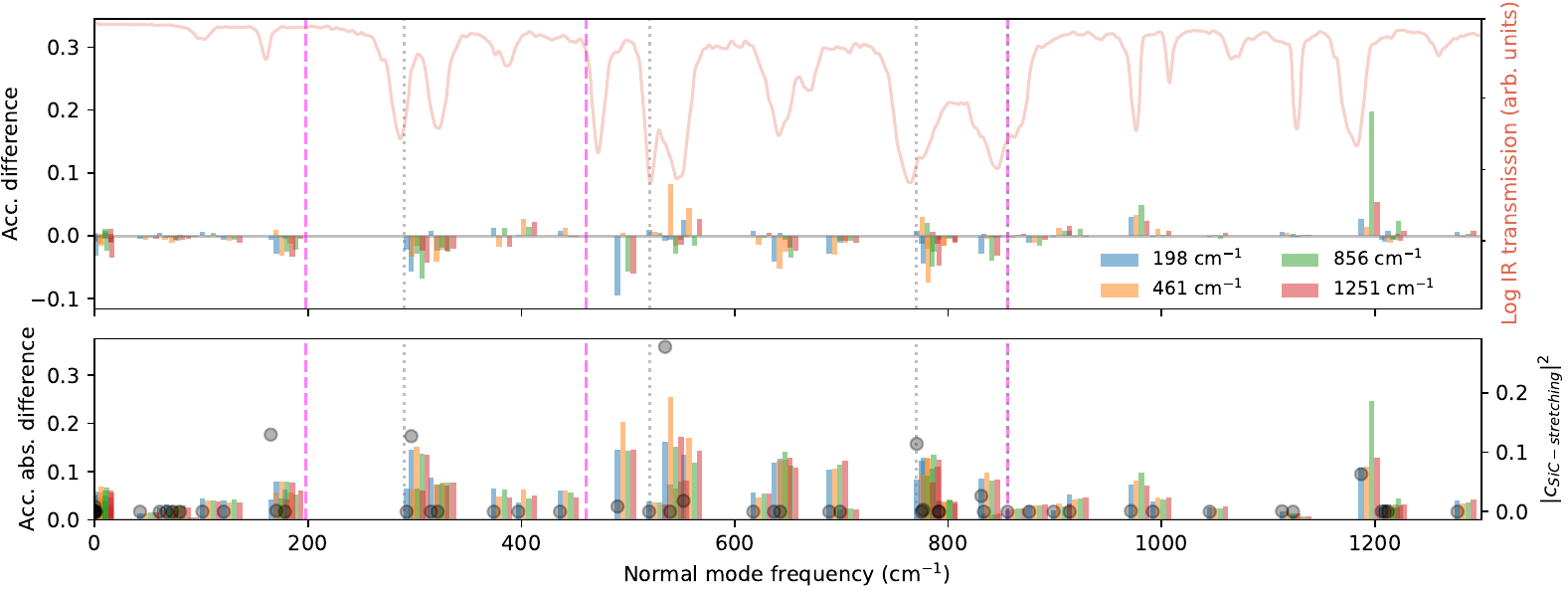}\\
    \caption{
    \textbf{Change in mode occupation due to coupling to cavity.}
    Accumulated (top) and accumulated absolute (bottom) difference in normal mode occupation for different cavity frequencies (given in cm$^{-1}$) with $g_0/\omega_c=1.132$ vs free space at 400~K.
    The corresponding differences in occupation and numerical details are presented in the SI. 
    We quantify the relative changes in cross-correlation between the IR spectrum and accumulated (absolute) difference in \autoref{tab:crosscorr}.
    \Cadd{Gray circles show the contribution of Si--C stretching to the vibrational normal mode (see text and SI \Cadd{Sec.~II F}).}
    Notice that we average here over a time-domain of 2~ps, and not 0.7~ps as in Ref.~\citenum{schafer2021shining}.
    }
    \label{fig:modediffs_accumulated}
\end{figure*}

\subsection{Vibrational Dynamics}

Let us shine a bit more light on the mechanistic differences between the chosen frequency domains that catalyze (461~cm$^{-1}$) or inhibit reactions without (198~cm$^{-1}$) or with (856~cm$^{-1}$) vibrational support.
\autoref{fig:modediffs_accumulated} illustrates the accumulated difference in normal mode occupation between a given cavity frequency and free space during the reaction, averaged over the full ensemble. 
The corresponding occupation differences are presented in the SI, where the overall structure for $\omega_c = 856$~cm$^{-1}$ is comparable to Ref.~\cite{schafer2021shining}.
Optically relevant domains around 300, 550, 770, and 1200~cm$^{-1}$ are noticeably affected more strongly \Cadd{when selecting the cavity frequencies $\omega_c=461$~cm$^{-1}$ and $\omega_c=856$~cm$^{-1}$}. 
\Cadd{Choosing a specific cavity frequency affects the vibrational modes in energetic proximity more strongly.
This effect is especially apparent for $\omega_c=856$~cm$^{-1}$ which involves the C=C stretching mode around 1200~cm$^{-1}$ (see green bars in \autoref{fig:modediffs_accumulated})}.
We quantify the changes in cross-correlation between the IR spectrum and the accumulated (absolute) difference (\autoref{fig:modediffs_accumulated} top (bottom)) in \autoref{tab:crosscorr} with the help of the relative difference between the cross-correlation for $\omega_c \in {198,~461,~856}$~cm$^{-1}$ and the high-frequency value $\omega_c=1251$~cm$^{-1}$.
The resulting change indicates how strongly differences in normal-mode occupation correlate with infrared activity.
\Cadd{Changes in the normal mode occupation for cavity frequencies with optical support, i.e., $\omega_c=461$~cm$^{-1}$ and $\omega_c=856$~cm$^{-1}$, feature larger correlation.}

\begin{table}[hb]
    \centering
    \caption{
    Relative difference of the cross-correlation function $\delta\delta C^{A(A)D}_{\omega_c} = \delta C^{A(A)D}_{\omega_c}/\delta C^{A(A)D}_{1251} - 1$, where we define $\delta C_{\omega_c} = \sum_i \vert \text{IR}(\omega_i) \cdot A(A)D_{\omega_c}(\omega_i) \vert / \sum_i \text{IR}(\omega_i)$, where $A(A)D$ represents the accumulated (absolute) difference.
    \Cadd{Only the} frequency domain illustrated in \autoref{fig:modediffs_accumulated} \Cadd{was utilized in the calculation of the cross-correlation function}.
    }
    \label{tab:crosscorr}
    \begin{tabular}{lrr}
    \toprule
    $ \omega_c$ (cm$^{-1}$) & $\delta\delta C^{AD}_{\omega_c}$ (\%) & $\delta\delta C^{AAD}_{\omega_c}$ (\%) \\
    \midrule
    198  &  +0.24 & +1.09 \\      
    461  &  +66.3 & +11.3 \\     
    856  &  +45.3 & +3.11 \\
    \bottomrule
    \end{tabular}
\end{table}

Overall stronger correlation for $\omega_c=461,~856$~cm$^{-1}$ suggests that the microscopic mechanism is more strongly characterized by redistribution of vibrational energy between optically active modes and is thus optically mediated, as suggested in  Ref.~\citenum{schafer2021shining}.
In other words, the cavity facilitates energy exchange between optically active modes\Cadd{, especially within an energy-window around the cavity frequency.} 
\Cadd{Since the Si--C bond is an essential ingredient in the reaction process, a stronger involvement of Si--C bond stretching (gray circles shown in \autoref{fig:modediffs_accumulated}) in the affected normal modes will result in a larger impact on the chemical reaction.
}
The vibrational analysis supports then also the previous hypothesis that chemical changes without support by optically active modes follow, to a certain degree, a different mechanism~\cite{schafer2021shining}.
\Cadd{Surprisingly, however, even though $\omega_c=461,~856$~cm$^{-1}$ seem to share a very similar mechanism based on the ML+MD analysis, their effect on the enthalpy is qualitatively different (catalysing vs inhibiting).
Albeit previous \textit{ab initio} calculations did not provide access to kinetic changes, the corresponding analysis provided no indication of a qualitative difference between $\omega_c=461$~cm$^{-1}$ and $\omega_c=856$~cm$^{-1}$~\cite{schafer2021shining}.
Let us reflect in the following section on the underlying approximation of our molecular dynamics simulations in order to understand this contradicting observation.}

\subsection{Limitations of simplified cavity molecular dynamics}
\label{sec:limitations}

The experimentally relevant domain around 856~cm$^{-1}$~\cite{thomas2016,thomas2020ground} is inhibited in our ML+MD calculations as well as in experiment and the recent \textit{ab initio} \gls{qedft} calculations with nuclear motion according to the Ehrenfest equations of motion~\cite{schafer2021shining}.
\Cadd{The existing information to changes in chemical rates is thus consistent at $\omega_c=856$~cm$^{-1}$.} 
On the other hand, the ML+MD calculations show enhanced rate constants \Cadd{at 461~cm$^{-1}$}, an effect that has not been observed in the QEDFT calculations.
A direct comparison of \autoref{fig:IRpowerSiC} to Ref.~\cite{schafer2021shining} is, however, problematic since the latter used an incomplete sampling with a strong bias towards the high-energy tail of the Boltzmann distribution.
In other words, the QEDFT results have been obtained at an effectively higher temperature.

\begin{figure}
    \centering
    \includegraphics[width=1.0\columnwidth]{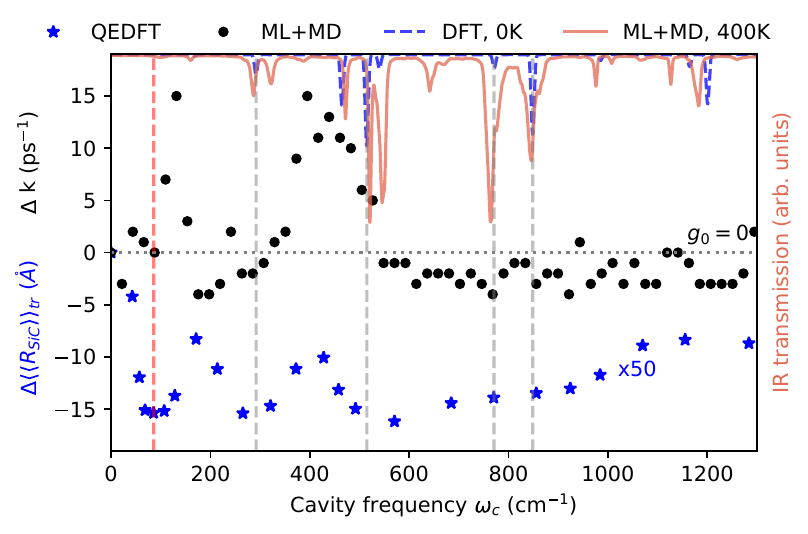}\\
    \includegraphics[width=1.0\columnwidth]{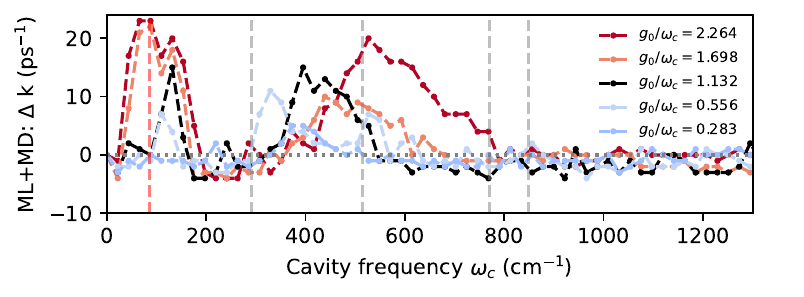}
    \caption{
    \Cadd{\textbf{Impact of dynamic electronic polarization.}}
    (Top)
    Black dots: Change in rate constant compared to free-space for the unidirectional reaction $\text{PTAF}^- \rightarrow \text{FtMeSi}+\text{PA}^{-}$ using the 30 initial configurations used in Ref.~\cite{schafer2021shining} and $g_0/\omega_c=1.132$, but propagated with our NEP based molecular dynamics calculator.\\ 
    Blue stars: Change in average Si--C distance obtained from the QEDFT calculations in Ref.~\cite{schafer2021shining}  amplified by a factor 50. Transmission spectrum obtained from DFT at 0K with harmonic approximation consistent with Ref.~\cite{schafer2021shining} (blue dashed), and using our NEP model and GPUMD at 400K NVE conditions (red solid). Vertical lines indicate characteristic features observed in Ref.~\cite{schafer2021shining}. We show in SI \Cadd{Sec.~II D} that Si--C distance and rate constant are clearly correlated, i.e., the qualitative trend of QEDFT and ML+MD can be set in relation.\\
    (Bottom) ML+MD calculations repeated for different fundamental light-matter coupling strength.}
    \label{fig:freqscanNatCom_combined}
\end{figure}

To allow for a more reliable comparison and shed light on the apparent discrepancy between the approaches, we recalculated the rate constant changes with our ML+MD approach for the same initial velocities as the QEDFT calculations. 
\autoref{fig:freqscanNatCom_combined} sets the newly obtained rate constants from our ML+MD calculations (black dots) in contrast with the average Si--C distances calculated with QEDFT (blue stars) and taken from Ref.~\cite{schafer2021shining}.
We demonstrate in SI \Cadd{Sec.~II D} that Si--C distance and rate constant are well correlated \Cadd{in this case}.

Ignoring the offset, a slight frequency shift, and the qualitatively different behaviour near $\omega_c=0$, the overall shapes of the reaction rate constant profiles shown in \autoref{fig:freqscanNatCom_combined} are consistent.
Given identical initial conditions, ML+MD and \gls{qedft} provide thus a similar \textit{profile} in the intermediate frequency domain but 
\Cadd{this profile is elevated into the catalysing domain for the ML+MD calculations.}
Looking back at \autoref{fig:IRpowerSiC}C, the major catalysing feature of the ML+MD calculations at $461$~cm$^{-1}$ is consistent between proper (\autoref{fig:IRpowerSiC}C) and incomplete sampling (\autoref{fig:freqscanNatCom_combined}).
Especially low-frequency features are considerably shifted and altered in strength, potentially originating from the quicker average reaction time when sampling from the high-energy tail of the Boltzmann distribution, which implies that the cavity has less time to influence the reaction.
We recall from \autoref{sec:kinetics}, that shorter reaction times emphasize dynamic contributions, calling the concept of a kinetic reaction rate further into question.
It is thus plausible that the overwhelming catalysing strength is partially an artifact and we therefore suggest to focus on the qualitative trend only.
This leaves but one question: 
Can we draw any conclusion about the mechanism of vibrational strong coupling from the observed discrepancy?

\subsubsection{\Cadd{Relevance of dynamic electronic polarization}} 

As we established at the beginning, effective nuclear forces are comprised of the adiabatic electronic Born-Oppenheimer forces and the dynamic optical forces mediated via dipolar changes induced by nuclear displacement.
This implies that the static polarization of the electric system due to the instantaneous cavity field as well as its non-adiabatic corrections, quantum nuclear, and quantum light-matter effects are absent -- as is the case in most available theoretical investigations for chemical reactivity affected by strong coupling.
Non-adiabatic electron-nuclear effects are expected to play a minor role as the electronic excited space is separated by about 3~eV at minimum and transition state.
While there has been recent discussions about the potential need to consider the full quantum light-matter interaction in model systems to recover resonant features in the cavity modified reactivity~\cite{lindoy2023quantum, fiechter2023quantum}, it remains up to debate if this is true for realistic systems under standard ambient conditions.
Such a question requires a nuanced discussion based on the specific system at hand and we expect that the answer will vary strongly between collective and single molecular coupling.

Lastly, and probably most importantly, \Cadd{nuclear motion will induce} strong optical fields \Cadd{which in turn} polarize the electronic system\Cadd{. This happens} similar to a static external potential $\propto g_0 q_c \boldsymbol\varepsilon_c\cdot\hat{\boldsymbol\mu}_e$, or via the self-polarization contributions $g_0^2/\omega_c (\boldsymbol\varepsilon_c\cdot\boldsymbol\mu_n)(\boldsymbol\varepsilon_c\cdot\hat{\boldsymbol\mu}_e)$ (and $g_0^2/(2\omega_c) (\boldsymbol\varepsilon_c\cdot\hat{\boldsymbol\mu}_e)^2$).
\Cadd{It is important to note that this polarization, although being instantaneous in the sense of the Born-Oppenheimer approximation, is inherently \textit{dynamic} due to its origin from the nuclear dynamics.
We will label this effect in the following \textit{dynamic} electronic polarization, to emphasize that the effect does not originate from the hybridization of the light-matter ground state nor from electronic transitions (non-adiabatic coupling elements).
This form of dynamic electronic polarization} can be formally incorporated via the cavity Born-Oppenheimer approximation~\cite{flick2017b, bonini2022}, where the photonic displacement $q_c$ is considered as an additional parametric variable and self-polarization contributions are added to the electronic structure calculations.
\Cadd{When included, the dynamic electronic polarization can lead to notable asymmetries in the vibrational polaritons~\cite{schafer2021shining,bonini2022} and increase the effective reaction barrier, thus reducing the chemical reactivity \cite{fischer2023,schnappinger2023cavity}.}
\Cadd{Our current ML+MD calculations are lacking the possibility to describe this dynamic electronic polarization and will thus tend towards a higher reactivity and more symmetric Rabi-splittings (see \autoref{fig:IRpowerSiC}~B).}

Consulting perturbation theory, cavity induced changes in the electronic ground-state energy scale approximately as $\propto g_0^2/\omega_c[1-\omega_c/(\Delta E_e + \omega_c)]$~\cite{pellegrini2015,schafer2021making}, where $\Delta E_e \gg \omega_c$ represents the electronic excitation energy for the dominant dipole transitions.
Perturbation theory should provide an adequate estimate \Cadd{for electronic changes} since the expansion parameter $\propto 1/\sqrt{V_c}$ is small and only $g_0/\omega_c$ takes appreciable values.
Fixing again $g_0/\omega_c=const. $ and noticing a larger dipole near the transition state (\Cadd{see Note}~\footnote{Notice in Fig.~3 of Ref.~\cite{riso2022molecular} the $\propto 1/\omega$ decay for fixed $V_c$, i.e., with fixed $ g_0/\omega_c$ one would obtain $g_0^2/\omega_c = g_0^2/\omega_c^2 \omega_c \propto \omega_c$}), an additional increase of the barrier  $\propto  \omega_c$ would be missing in our MD calculations compared to the QEDFT calculations.
As a result, \Cadd{our ML+MD calculations provide sensible values near $\omega_c=0$ and for large frequencies, but have the tendency to be overly reactive in the intermediate domain as they lack an additional suppression from dynamic electronic polarization.}

\Cadd{
Larger cavity frequencies $\omega_c$ are driven resonantly only by vibrational modes of comparable frequency.
If such a vibration contributes to the reaction, i.e., whether it will be noticeably excited during the reaction, depends largely on its Si--C contribution.
If the cavity can exert notable effects on a reaction event will therefore depend on three time scales: (i) The strongly occupied reactive modes of frequency $\omega_\text{SiC}$, (ii) the cavity frequency $\omega_c$, and (iii) the frequency  with which the PES is modulated $\omega_\text{modPES}$ (dynamic electronic polarization) as consequence of the nuclear dynamics.
Since $\omega_\text{modPES}$ depends partially on the optical mode $q_c$, which oscillates with $\omega_{c}$, it seems intuitive that the largest effect of dynamic electronic polarization on the reactive modes $\omega_\text{SiC}$ can be expected when all time scales are comparable.
As emphasized by \autoref{fig:IRpowerSiC}B and \autoref{fig:modediffs_accumulated}, most of the optically active modes with relevant Si--C contribution are located below 840~cm$^{-1}$.
The \textit{dynamic} contribution of electronic polarization will then play a decreasing role at higher frequencies.
}
How the precise interplay between nuclear motion, strongly coupled cavity, and cavity-modulated electronic polarization affects the reactivity goes beyond the scope of this work.
\Cadd{Our results emphasize the role of dynamic electronic polarization but also indicate that it is unlikely to be the only relevant contribution.}
\Cadd{Ample experimental work, however, demonstrated changes in solute-solvent interaction~\cite{doi:10.1021/jacs.3c02260,singh2023solvent} 
under vibrational strong coupling which indicates the involvement of dispersive interactions mediated by electronic polarization.}

\subsubsection{\Cadd{Additional considerations}} 

\begin{figure}
    \centering
    \includegraphics[width=1.0\columnwidth]{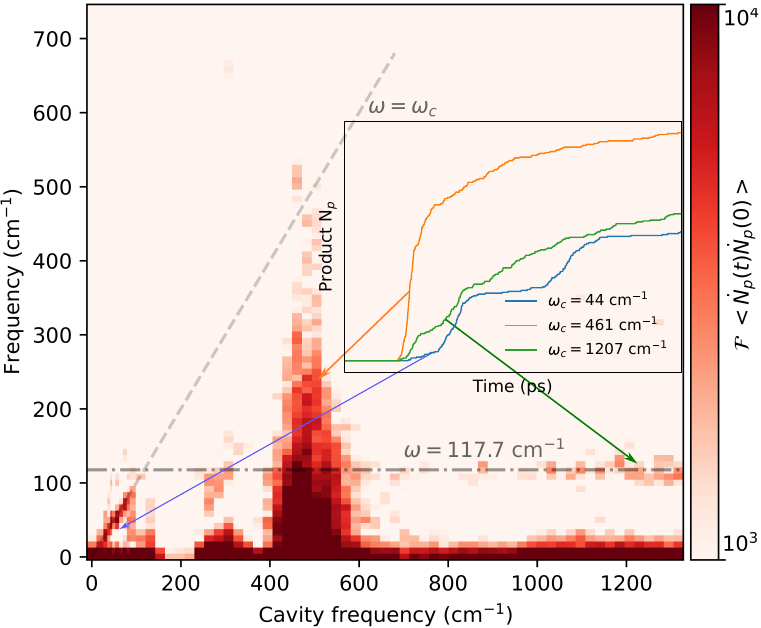}
    \caption{\Cadd{\textbf{Frequency dependence of reaction bursts.}} Fourier transform of the autocorrelation function of changes in the number of products $\vert \mathcal{F} \langle\dot{N}_p(t) \dot{N}_p(0)\rangle \vert$ at 400 K and $g_0/\omega_c=1.132$ where ${N}_p$ is the number of products.
    The inset shows the time-dependent accumulation of products at low frequencies, in the strongly catalysing domain, and at high cavity frequencies where the 117~cm$^{-1}$ mode attains a pronounced role during the reaction (see text).}
    \label{fig:fft_autocorrNproductT}
\end{figure}

Let us consider the analogy of our MD calculations as the self-consistently driven dynamic of a ball on a high-dimensional energy surface.
If we intend to cross a specific barrier, but let the cavity periodically remove kinetic energy before inserting it back at a later time, we can imagine that the likelihood to cross the barrier is modulated by the frequency with which the cavity is oscillating.
We can indeed observe such an effect at low cavity frequencies where reaction events appear in bursts that are related to the cavity frequency.
\autoref{fig:fft_autocorrNproductT} shows the Fourier transform of the auto-correlation function of the change in product of the S$_N$2 reaction.
The linear dispersion at low frequencies clearly shows that the bursts in reaction events are correlated with the cavity frequency, a feature that is absent in the QEDFT calculations of Ref.~\cite{schafer2021shining}.
This trend continues up to a bending mode at 117~cm$^{-1}$, which contributes to the necessary rearrangement of the methyl groups. Bending modes have recently been identified as a relevant component in cavity-enhanced charge transfer \cite{kumar2023extraordinary}, which might suggest that the observed interplay between linear dispersion and bending mode could be of wider relevance.
Surely, this simplified picture of modulated reactivity can only hold if vibrational energy redistribution into other modes is unlikely -- a condition that is rarely fulfilled and explains why the effect disappears quickly. 

It seems intuitive to assume that the here observed discrepancy, supposedly scaling with $g_0^2$, should only matter for sizeable coupling strengths, and the change in rate constant should approach the cavity-free reference value \Cadd{monotonously} for decreasing coupling.
However, this would either imply that the ML+MD calculation would need to qualitatively change from catalysis to inhibition while approaching $k_{g_0=0}$, which is not observed in \autoref{fig:freqscanNatCom_combined} (bottom), or that the QEDFT calculations should become catalysing during this process, for which there was no indication in Ref.~\citenum{schafer2021shining}.
One might thus draw the conclusion that the \Cadd{dynamic electronic polarization,} missing in ML+MD\Cadd{, is required for the} qualitative prediction of chemical changes via single-molecule vibrational strong coupling \Cadd{at most frequencies and for all coupling values}.
\Cadd{Furthermore, such features might even play a role in collectively coupled systems where individual molecules can exhibit large dynamic electronic polarization in response to collective vibrational dynamics~\cite{sidler2023unraveling,schnappinger2023cavity}.}

\Cadd{
SI Sec.~II~H includes the same investigation performed in this section using a second NEP model based on DFT data calculated with a smaller electronic basis-set. The overall trend up to 700~fs (as in Ref.~\citenum{schafer2021shining}) is consistent with the here presented ML+MD calculations. Since both ML+MD investigations show a consistent qualitative deviation from the QEDFT calculations, they both support the argument of a dynamic electronic polarization.
In addition, the smaller basis-set results in a smaller reaction barrier which reduces the significance of the catalysed trajectories at longer times.
The most characteristic feature remaining after 2~ps is a strong inhibition around 200~cm$^{-1}$, consistent with \autoref{fig:IRpowerSiC}C.
This emphasizes that estimates of kinetic quantities require sufficient sampling.
}

Sampling, calculating, and learning potentials for every frequency and coupling, in order to obtain the full cavity Born-Oppenheimer surface, is in principle possible.
If we intend to widely scan those parameters or would like to include more than a single mode, it becomes, however, practically unfeasible.
A (self-consistent) perturbative correction~\cite{bonini2022,10.1063/5.0153293,fischer2023} to the electronic quantities might suffice to suppress some of the problematic features.
Such a correction could be constructed via static electronic polarizabilities, which can be predicted with an additional neural network~\cite{xu2023dipole}, to correct the electronic energies and dipole moments according to the adiabatic cavity field.
We envision to combine such a corrected treatment with a full GPU implementation of our approach, which would dramatically reduce the required computational time and pave the way towards explicit ensembles, solvents, and accurate estimates of the chemical kinetics.

\section{Conclusion}\label{sec:conclusion}

Recent years have seen a rapid development of theoretical models for the description of strong light-matter coupling, and polaritonic chemistry in particular.
Most theoretical work focuses on model systems, which is natural given the young age of the field, and yet poses a major challenge as overly simplified models are unable to truly connect to experiment.
Here, we illustrate a combination of \textit{ab initio} trained machine learning models and modular cavity molecular dynamics that aims to describe realistic molecules in realistic optical environments.
This work allows, for the first time, to directly relate theoretical predictions to the experimentally measured changes in chemical kinetics.

We describe theoretically the appearance of single-molecule strong coupling and its influence on the Si--C bond for the experimentally investigated S$_N$2 reaction~\cite{thomas2016}.
A clear frequency dependence of the rate constant, a critical aspect of polaritonic chemistry, is observed and translates into changes in enthalpy and entropy that are consistent with experimental observations.
Interestingly, we observe inhibiting and \emph{catalysing} effects for the same reaction, the latter of which stand in contrast to previous \textit{ab initio} calculations \cite{schafer2021shining}. 
In total, three different regimes can be identified that are set apart by differences in the kinetic changes.
\emph{(i)} A strongly inhibiting effect without clear vibrational contribution results in a strong increase in enthalpy but relatively small increase in entropy.
\emph{(ii)} Vibrationally supported catalysis predominantly increases the entropy and only slightly lowers the enthalpy, which results effectively in a simple shift in the Eyring plot, but we emphasize, that the shorter reaction time likely results in an overestimation of this effect.
\emph{(iii)} Vibrationally supported inhibition raises the enthalpy and results in the strongest increase in entropy.
The latter observation is qualitatively consistent with experiment and suggests a slight change in the chemical character of the reaction.
Vibrationally supported \Cadd{rate changes} \emph{(ii-iii)} are accompanied by a more pronounced change in normal-mode occupation in optically active domains, suggesting that the microscopic mechanism is caused by a stronger interplay of optically active modes via the cavity, i.e., by cavity mediated changes in the redistribution of vibrational energy.

The discrepancy with \textit{ab initio} calculations, although sharing comparable patterns when scanning the cavity frequency and comparing identical ensembles, suggests that \Cadd{dynamic changes in} electronic polarization \Cadd{induced by nuclear motion and mediated by the cavity play} a considerable role at the selected coupling strength.
This might explain why many simplified molecular dynamics simulations aiming to understand vibrational strong coupling have been able to capture some frequency dependence but often showed strong detuning, or simultaneous catalysing and inhibiting features for the same reaction, which, to best of our knowledge, is in conflict with current experimental work.
Our work demonstrates therefore the importance of \textit{ab initio} QED in the future of polaritonic chemistry and emphasizes the significance of theory that is tailored to describe experimentally relevant reactions.
Future development will focus on (self-consistent) perturbative corrections and the full transfer of the established framework to graphical processor units (GPUs), thus providing access to realistic (optical) environments and explicit description of solute-solvent ensembles.
The latter is essential for investigating potential modification of solvation dynamics induced by strong coupling~\cite{castagnola2023collective,doi:10.1021/jacs.3c02260,singh2023solvent}.

Polaritonic chemistry remains an equally fascinating and puzzling domain of research.
While major questions are yet to be answered~\cite{garcia2021manipulating, simpkins2021mode, sidler2021perspective, mandal2022theoretical}, especially the connection between local chemistry and collective coupling as well as the interplay with solvation, the continuous growth of theoretical methodology and additional experiments draw an optimistic picture for the future of polaritonics.
Our work adds to this a new facet and a clear perspective for possible future development, providing valuable insight that can be experimentally validated.

\section*{Supporting Information}
Information on training procedure and performance estimates for the NEP models, Numerical and Simulation details, Derivation and Implementation of optical forces, NVT calculations with corresponding estimates for enthalpy and entropy, Details for kinetic estimates from NVE calculations (including Arrhenius plots), Details and results for rate and Si--C distance estimates for consistency check, Details for normal mode projections and Si--C stretching contributions, Units, A second evaluation of the consistency check based on a different NEP model trained from DFT calculations using the smaller 6-31G* basis, Extensive information and data on the calculation of vibrational frequencies (intermediate and transition state).

\section*{Acknowledgments}
We thank Zheyong Fan, Magnus Rahm, Stefano Corni, and G\"oran Johansson for insightful discussions.
C.S. acknowledges support from the Swedish Research Council through Grant No. 2016-06059 and funding from the Horizon Europe research and innovation program of the European Union under the Marie Sk{\l}odowska-Curie grant agreement no.\ 101065117.
J.F., E.L., and P.E. acknowledge funding from the Knut and Alice Wallenberg foundation through Grant No. 2019.0140, funding from the Swedish Research Council through Grant No. 2020-04935 as well as the Swedish Foundation for Strategic Research via the SwedNESS graduate school (GSn15-0008).
The computations were enabled by resources provided by the National Academic Infrastructure for Supercomputing in Sweden (NAISS) at NSC, PDC, and C3SE partially funded by the Swedish Research Council through grant agreement no. 2022-06725.

Partially funded by the European Union.
Views and opinions expressed are, however, those of the author(s) only and do not necessarily reflect those of the European Union or REA.
Neither the European Union nor the granting authority can be held responsible for them.

\section*{Data availability}
\Cadd{Training data, inputs, and final NEP model are available via Zenodo \url{https://doi.org/10.5281/zenodo.10255268}.}


\bibliography{references}
\end{document}


\title{Supplementary Information to Machine Learning for Polaritonic Chemistry: Accessing chemical kinetics}

 \author{Christian Sch\"afer}
 \email[Electronic address:\;]{christian.schaefer.physics@gmail.com}
 \affiliation{
 Department of Physics, Chalmers University of Technology, 412 96 G\"oteborg, Sweden}
 \affiliation{
  Department of Microtechnology and Nanoscience, MC2, Chalmers University of Technology, 412 96 G\"oteborg, Sweden}
 \author{Jakub Fojt}
 \affiliation{
 Department of Physics, Chalmers University of Technology, 412 96 G\"oteborg, Sweden}

 \author{Eric Lindgren}
 \affiliation{
 Department of Physics, Chalmers University of Technology, 412 96 G\"oteborg, Sweden}

  \author{Paul Erhart}
  \affiliation{
  Department of Physics, Chalmers University of Technology, 412 96 G\"oteborg, Sweden}

\date{\today}

\maketitle

\section{Supplementary Methods}

\subsection{Preparing the training set}
\label{sec:train}
We use an active learning approach in this work. An initial training set includes unrelaxed structures from the potential energy surface published in Ref.~\cite{schafer2021shining} and additional structures generated by rattling the minimum energy PTAF$^-$ structure. 
We then used the ORCA code version 5.0~\cite{orca2018} (PBE, 6-31G$^*$ basis) to obtain energies, forces, and dipoles.
We trained a first version of the NEP model, using the NEP-3 potentials in GPUMD~\cite{fan2022gpumd}. 
With this first generation model, we performed molecular dynamic simulations and selected trajectories with bad performance, adding those structures to the training set and repeating the procedure for a total of 7 generations.
It should be noted that GPUMD has been undergoing changes since those initial attempts.

The comparably small 6-31G$^*$ basis provided reasonable energies but showed limited reliability for the dipole moments. We prepared a new dataset that included all rattling and molecular dynamics structures as well as a randomly sampled set of structured from the potential energy surface. We performed DFT PBE def2-TZVP calculations (using tight SCF convergence) to generate in total 20170 structures to train the final dipole NEP model with the following parameters.
\begin{verbatim}
mode     1
version  4
type     4 Si F C H
cutoff   8 6
n_max    15 8
neuron   80
batch    500000
generation  500000
\end{verbatim}
The energy/force model has been trained on the same dataset with the input parameters.
\begin{verbatim}
version  4
type  4 Si F C H
cutoff  8 4
n_max  8 6
l_max  4 0
lambda_1  0.1
lambda_2  0.1
lambda_e  1
lambda_f  3
lambda_v  0
neuron  40
batch  300000
generation  500000
\end{verbatim}
Both models are sufficiently converged and their performance is illustrated in the following sections.

\subsection{\Cadd{Initial performance estimates}}

\begin{figure}
    \centering
    \includegraphics[width=1.0\columnwidth]{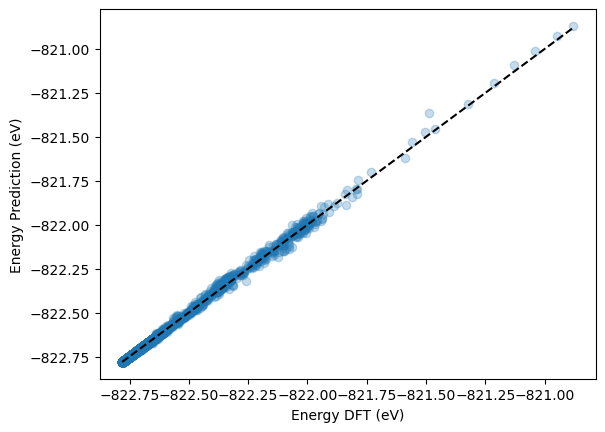}
    \caption{Simple scatter-plot for the energy model. We use here all training structures but show additional validation checks in the following sections.}
    \label{fig:energy_scatter}
\end{figure}

Let us start our model evaluation with the simple scatter plot Fig.~\ref{fig:energy_scatter} for the expected and predicted energies. The model is well converged within its training set. Additional tests for an independent test-set follow later.

\begin{figure}
    \centering
    \includegraphics[width=1.0\columnwidth]{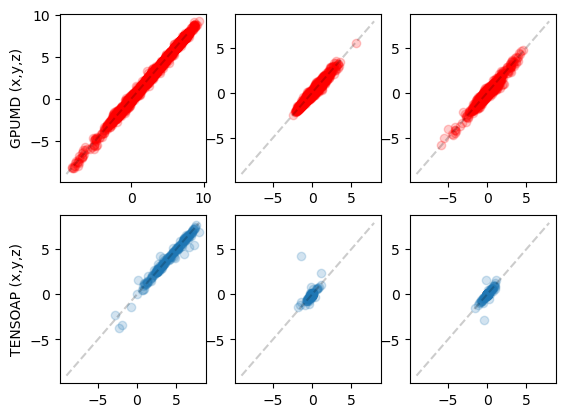}
    \caption{Quality assessment of the prediction for the x,y, and z component (left to right) of the GPUMD and TENSOAP model. We use all 20170 structures for the GPUMD scatter plot. The SA-GPR model of TENSOAP used 800 structures for training and 193 for the here shown scatter plot.}
    \label{fig:dipole_tensoap_bench}
\end{figure}

We compared our dipole model against the established symmetry adapted Gaussian Process Regression (SA-GPR) employed by TENSOAP \cite{PhysRevLett.120.036002} for which we randomly selected a set of 993 structures. Fig.~\ref{fig:dipole_tensoap_bench} shows dipole predictions for GPUMD and TENSOAP. SA-GPR is considerably slower and practical application for molecular dynamics is limited as it requires at each step to build kernel elements between the test and training set. That said, they require typically less data. For this reason, we decided to use less data for SA-GPR which keeps the training time and memory requirements low and allows for somewhat comparable evaluation times, would one perform molecular dynamics with both models.

Our NEP model performs overall well and is quicker to evaluate than SA-GPR, making it the more convenient choice for our purpose. This short discussion is not suited to make a general claim about the superiority of one of the models but merely serves as sanity check among the existing approaches. 


\subsection{Validation of the NEP models compared to ab initio calculations}

\begin{figure}
    \centering
    \includegraphics{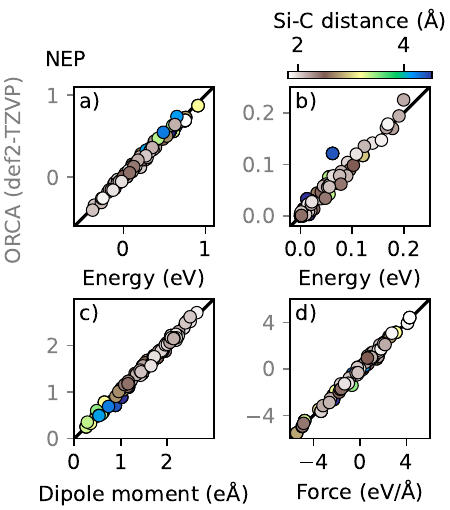}
    \caption{Parity plots for our NEP model versus ab initio calculations performed with ORCA and the def2-TVZP basis set.
    Comparison of (a) potential energies of the electronic system (we have subtracted \qty{-22214}{\electronvolt} from the values), (b) kinetic energies of the electronic system, (c) dipole moments and (d) the (electronic + cavity) force acting on the Si-C bond projected on the bond vector.
    }
    \label{fig:parity_plots_NEP_ORCA}
\end{figure}

To validate our NEP model against data outside the training set, we select 99 configurations for various trajectories, at various times, from our MD simulations performed in SI Sec.~\ref{sec:natcomcheck}. This provides an estimate for the quality of the model in general and the specific reliability of our reruns from which we draw the conclusion that electronic polarization might play a more relevant role than currently thought. 
For each configuration we evaluate the potential and kinetic energies of the electronic system, the dipole, and the total force (both electronic and cavity contributions) on the Si-C bond using the NEP model, and using ORCA (\autoref{fig:parity_plots_NEP_ORCA}). 
NEP model and ORCA are overall in good agreement, and we can expect that our NEP model is able to accurately reproduce the trajectories that one would obtain from \textit{ab initio} MD.
There is one outlier among the data for the kinetic energy featuring a large Si-C bond distance (bond is broken).
ML models are excellent at interpolation but configurations that push beyond the tightly sampled domain will suffer from reduced accuracy. However, those configurations do not play any role in our analysis since the bond is considered as broken once it crosses a Si--C distance of 3.5\AA, i.e., any data generated for larger values do not enter into our analysis.

\begin{figure}
    \centering
    \includegraphics{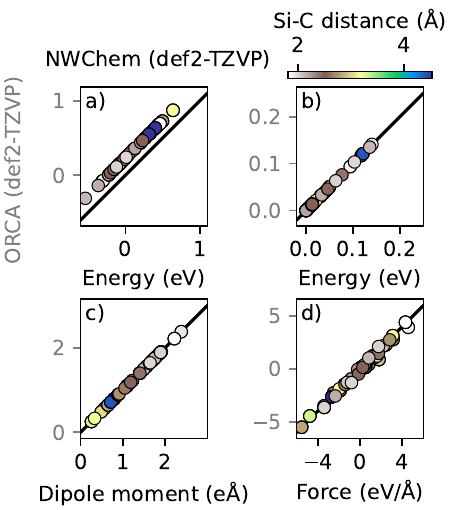}
    \caption{Parity plots for ab initio calculations performed with ORCA versus NWChem.
    The def2-TVZP basis was used in both cases.
    Comparison of (a) potential energies of the electronic system (we have subtracted \qty{-22214}{\electronvolt} from the values), (b) kinetic energies of the electronic system, (c) dipole moments and (d) the (electronic + cavity) force acting on the Si-C bond projected on the bond vector.}
    \label{fig:parity_plots_ORCA_NWCHEM}
\end{figure}

\begin{figure}
    \centering
    \includegraphics{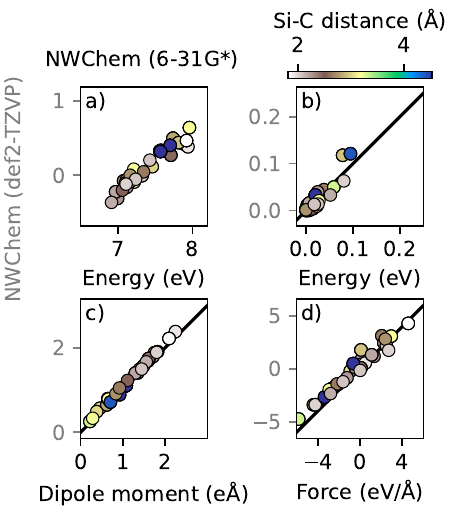}
    \caption{Parity plots for ab initio calculations using the def2-TVZP versus 6-31G* basis sets.
    Calculations were performed using NWChem.
    Comparison of (a) potential energies of the electronic system (we have subtracted \qty{-22214}{\electronvolt} from the values), (b) kinetic energies of the electronic system, (c) dipole moments and (d) the (electronic + cavity) force acting on the Si-C bond projected on the bond vector.}
    \label{fig:parity_plots_NWCHEM_BASIS}
\end{figure}

We perform additional consistency checks between the DFT codes ORCA and NWChem with the def2-TVZP basis set (\autoref{fig:parity_plots_ORCA_NWCHEM}) and for two different basis sets def2-TVZP and 6-31G* within NWChem (\autoref{fig:parity_plots_NWCHEM_BASIS}). ORCA and NWChem are in good agreement, besides a small and irrelevant constant shift in the potential energy. However, the difference between the basis sets is larger than the difference between NEP model and reference calculations, which provides further evidence for the quality of our model in the relevant domain.

\subsection{Numerical Details}

\begin{figure}
    \centering
    \includegraphics[width=1.0\columnwidth]{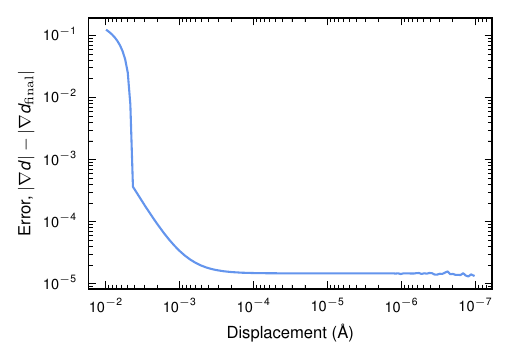}
    \caption{Finite-difference error as a function of finite difference displacement size. The error is calculated as the difference between the norm of the dipole gradient for a certain displacement and for the smallest displacement \qty{e-8}{\angstrom}. Note that the error plateaus for displacements smaller than \qty{e-4}{\angstrom}, which could be an effect of reaching the limit of the accuracy for the dipole predictions from the model.}
    \label{fig:fderror}
\end{figure}

All ASE calculations use the Velocity Verlet propagator with a time-step of 0.1~fs. We obtained the Jacobian of the dipole moment with a 2nd-order central-difference approximation using displacements $h=10^{-4}$~\AA, implemented in the Python package calorine \cite{lindgren2023calorine}.  This value is located at the beginning of a stable plateau illustrated in Fig.~\ref{fig:fderror}, i.e., smaller steps are not useful.

\begin{figure}
    \centering
    \includegraphics[width=1.0\columnwidth]{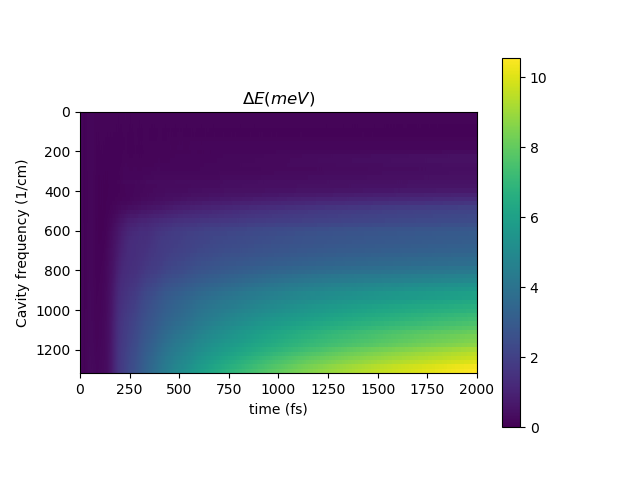}
    \caption{Average energy loss in trajectory bundle over time for 400 K and $g/\omega=1.132$.}
    \label{fig:energyloss}
\end{figure}

The ensemble averaged energy loss during propagation is given in Fig.~\ref{fig:energyloss}. It increases with increasing frequency since the ratio $g/\omega$ is kept constant and a larger cavity frequency leads thus to stronger cavity induced forces which in turn increase the accumulated error due to the finite-difference approximation. We plan to extend GPUMD with analytic derivatives in the future which would entirely mitigate this error.

\Cadd{
Our NVE calculations set a temperature by sampling initial velocities from a Boltzmann distribution and removing the center of mass momentum.
Ref.~\cite{schafer2021shining} showed that considering a solvent resulted in notable changes of the relative infrared activity of vibrational excitations but the good agreement in enthalpy (using NVT conditions) suggests that the effect on the reaction is small enough to draw relevant conclusions from our investigations.
}

\section{Supplementary Information}

\subsection{Obtaining total forces from electronic forces and dipoles}
\label{sec:interfacing-cavity}

Splitting the Hamiltonian in matter $\hat{H}_{0}$ and light-matter component
\begin{align}
    \hat{H} &= \hat{H}_{0} + \frac{1}{2}[\hat{p}^2 + \omega_c^2 (\hat{q}-\frac{1}{\sqrt{\varepsilon_0 V_c}}\varepsilon_c \cdot \hat{\boldsymbol{\mu}} /\omega_c)^2]
\end{align}
and introducing the simplified classical limit $ \hat{q},\hat{p},\hat{\boldsymbol{\mu}} \rightarrow q, p, \boldsymbol{\mu} $, we obtain the classical Hamilton function
\begin{align}
    \mathcal{H}_{LM} = \frac{1}{2}[p^2 + \omega_c^2 (q-\frac{1}{\sqrt{\varepsilon_0 V_c}}\varepsilon_c \cdot \boldsymbol{\mu} /\omega_c)^2].
\end{align}
The careful reader will notice that the classical nuclear limit does not strictly imply $\hat{\boldsymbol{\mu}} \rightarrow \boldsymbol{\mu}$ since the total dipole moment includes electronic and nuclear contributions. The electronic remainder, especially $(\varepsilon_c \cdot \hat{\boldsymbol{\mu}}_e)^2 + \varepsilon_c \cdot \hat{\boldsymbol{\mu}}_e \varepsilon_c \cdot \boldsymbol{\mu}_n$, polarizes the electronic system and thus influences the nuclear forces. We discuss the potential consequences of this subtlety and the formally correct treatment in terms of the cavity Born-Oppenheimer~\cite{flick2017b} in further detail in the main text.

Following classical Hamilton mechanics for the canonical displacement mode of the cavity oscillator
\begin{align}
    \partial_t p &= \{p,\mathcal{H}_{LM}\} = - \frac{\partial \mathcal{H}}{\partial q}\\
    &= -\omega_c^2 q + \omega_c \frac{1}{\sqrt{\varepsilon_0 V_c}}\varepsilon_c \cdot \boldsymbol{\mu}
\end{align}
provides, due to the equivalence of kinetic and canonical momentum in Power-Zienau-Wooley gauge $\partial_t q = \{q,\mathcal{H}_{LM}\} = \frac{\partial \mathcal{H}}{\partial p} = p$, the mode-resolved Maxwell equation
\begin{align}
    (\partial_t^2 + \omega_c^2) q(t) = \omega_c \frac{1}{\sqrt{\varepsilon_0 V_c}}\varepsilon_c \cdot \boldsymbol{\mu}(t),
\end{align}
solved by Green's function $G(t-t') = \frac{\sin(\omega_c(t-t'))}{\omega_c}$.

Enforcing zero initial cavity momentum $p(0) = 0$, the cavity mode displacement $q(t)$ depends then on the time-evolution of the molecular dipole moment through
\begin{align}
    \label{eq:q-time-dependent}
    q(t) =
         q(0) \cos(\omega_c t)
         + \int_0^t \frac{\varepsilon_c\cdot\boldsymbol{\mu}(t')}{\sqrt{\varepsilon_0 V_c}}  \sin(\omega_c (t - t'))\:\mathrm{d}t'.
\end{align}
\Cadd{The initial mode displacement is chosen such that the initial optical force is zero, i.e., $q_0=\frac{1}{\omega_c}\frac{1}{\sqrt{\varepsilon_0 V_c}}\varepsilon_c \cdot \boldsymbol{\mu}_0$. }
To avoid having to store the entire time-evolution of $\boldsymbol{\mu}$ during MD, we decouple $t$ and $t'$ in \autoref{eq:q-time-dependent}
\begin{align}
    q(t) &=
         q(0) \cos(\omega_c t) + \sin(\omega_c t)C(t) - \cos(\omega_c t)S(t) \\
    C(t) &= \int_0^t \frac{\varepsilon_c\cdot\boldsymbol{\mu}(t')}{\sqrt{\varepsilon_0 V_c}}  \cos(\omega_c t')\:\mathrm{d}t' \\
    S(t) &= \int_0^t \frac{\varepsilon_c\cdot\boldsymbol{\mu}(t')}{\sqrt{\varepsilon_0 V_c}}  \sin(\omega_c t')\:\mathrm{d}t',
\end{align}
and update the integrals during MD as
\begin{align}
    C(t + \Delta t) &= C(t) + \int_t^{t+\Delta t} \frac{\varepsilon_c\cdot\boldsymbol{\mu}(t')}{\sqrt{\varepsilon_0 V_c}}  \cos(\omega_c t')\:\mathrm{d}t' \\
    S(t + \Delta t) &= S(t) + \int_t^{t+\Delta t} \frac{\varepsilon_c\cdot\boldsymbol{\mu}(t')}{\sqrt{\varepsilon_0 V_c}}  \sin(\omega_c t')\:\mathrm{d}t',
\end{align}
where the trapezoidal rule is used to approximate the integrals.

We emphasise that $\boldsymbol{\mu}$ and $\boldsymbol{F}_\text{PES}$ can be obtained either directly from DFT, or our NEP models, but only the latter is computationally tractable for MD simulations.

\subsection{NVT reference calculation}

Theoretically predicted rates require sufficient time for thermalization to reach a statistically meaningful distribution near an equilibrium state of the system. This requires long propagation times and suggests the use of NVT conditions. Both aspects are problematic in calculations involving the cavity for two major reasons. First, the interplay between thermostat and cavity might give rise to spurious features that misguide our interpretation. Second, observing an appreciable number of reactions under such conditions requires long propagation times. The latter is not an issue for calculations on the GPU, but the current CPU based cavity calculator is certainly limited in this aspect.

\begin{figure}
    \centering
    \includegraphics[width=1.0\columnwidth]{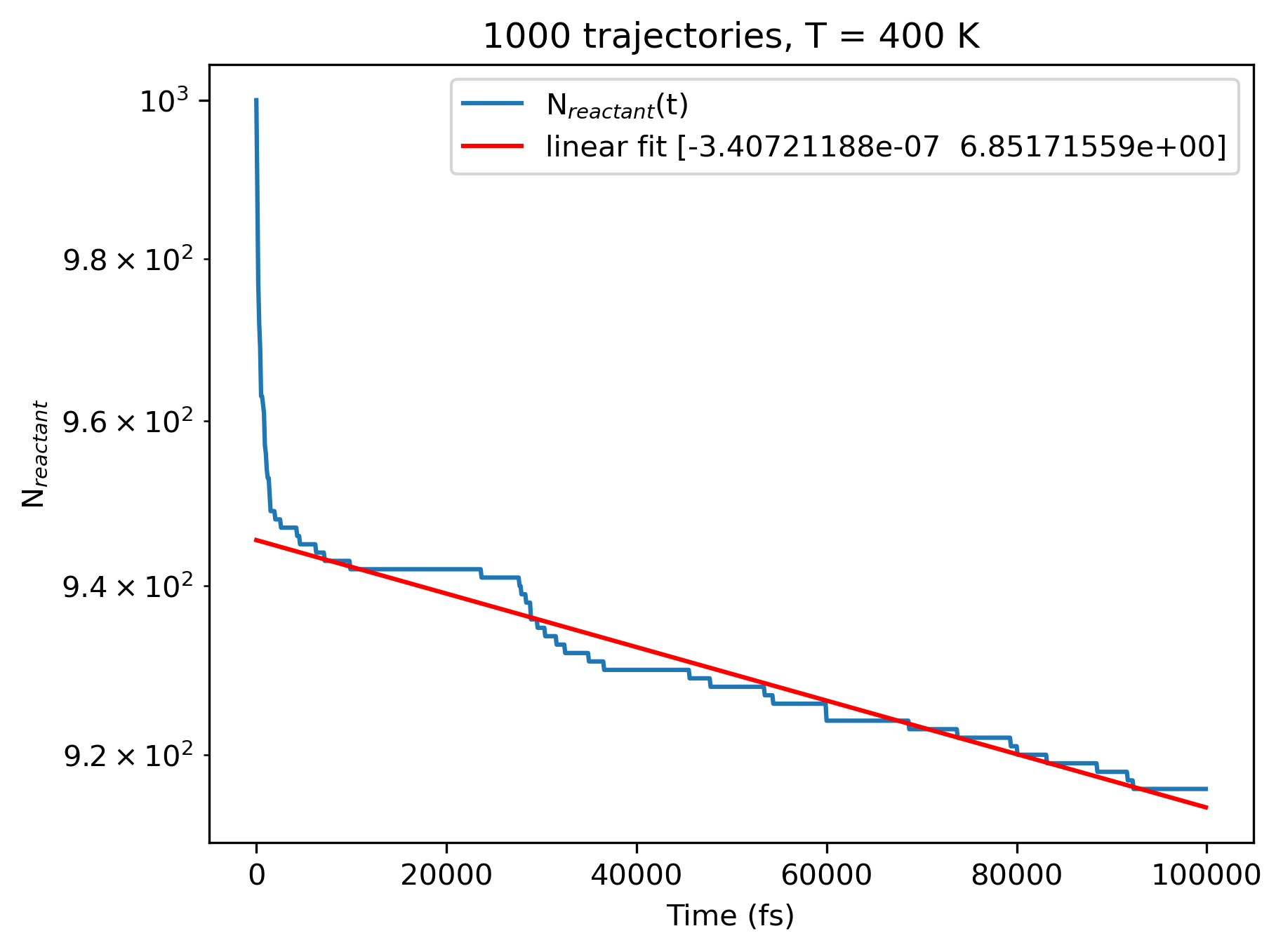}
    \caption{
    Log-plot of the number of reactant molecules $PTAF^-$ vs time and linear fit to the domain after initial equilibration. NVT conditions with 400K were enforces using the GPUMD internal Nos\'e-Hoover chain thermostat with relaxation time-value of 100. A time-step of 0.1~fs was used.
    }
    \label{fig:Nreactant_NVT}
\end{figure}

\begin{figure}
    \centering
    \includegraphics[width=1.0\columnwidth]{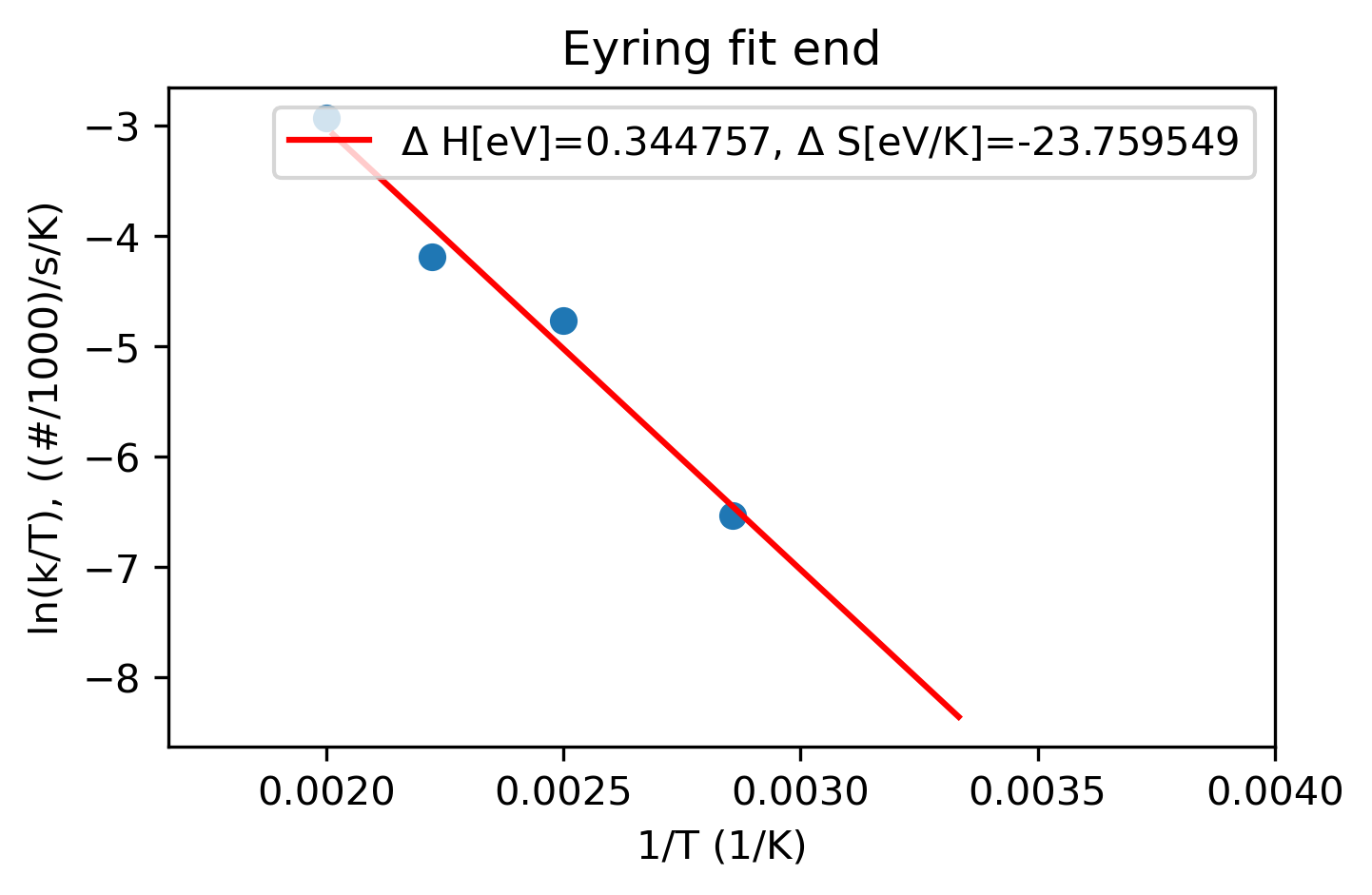}
    \caption{
    Eyring plot and fit for the unidirectional reaction $PTAF^- \rightarrow FtMeSi+PA^{-}$ obtained under NVT conditions according to Fig.~\ref{fig:Nreactant_NVT}. Fit to the thermalized reaction dynamics. The latter predicts an enthalpic barrier of $\Delta H^\ddagger = 0.345$~eV, which is consistent with the experimentally measured  $\Delta H^\ddagger = 35\pm4$~kJ/mol.~\cite{thomas2020ground}}
    \label{fig:Eyring_NVT}
\end{figure}

Fig.~\ref{fig:Nreactant_NVT} presents an example of the number of reactant molecules over time at 400K. We observe two characteristic domains, an initial burst of reactions, and a second domain for the properly thermalized reactant. We extract the rates as linear fits to the latter and calculate a transition-state enthalpy of $\Delta H^\ddagger = 0.345$~eV from the Eyring plot in Fig.~\ref{fig:Eyring_NVT}, which is in agreement with experimental estimates $\Delta H^\ddagger = 35\pm4$~kJ/mol.~\cite{thomas2020ground} It should be noted that we only estimate the Si-C breaking reaction step here, as the correct fluoride attacking and final protonation steps would require an explicit treatment of the solvent.

\subsection{Rate calculations and thermodynamics from NVE}

A reaction event took place when the Si--C bond is stretched beyond a value of 3.5~\AA. This is located about 0.5~\AA~ behind the transition state, allowing us to account for recrossing events in a simplified manner. Once a trajectory showed such a reaction event, it is considered as product in the following. The rate is calculated as number of products after 2~ps. Our rate constant is therefore that of a unidirectional reaction towards the product and does not correspond to the equilibrium rate. Rate constant is calculated for different temperatures, recall that the initial velocities are samples from the Maxwell-Boltzmann distribution, and plotted in the Eyring plots. The linear fits with a first-order polynomial provides estimates for enthalpy and entropy for this reaction process. The thermodynamic quantities are presented as absolute difference to the cavity-free calculations. 

\begin{figure}
    \centering
    \includegraphics[width=1.0\columnwidth]{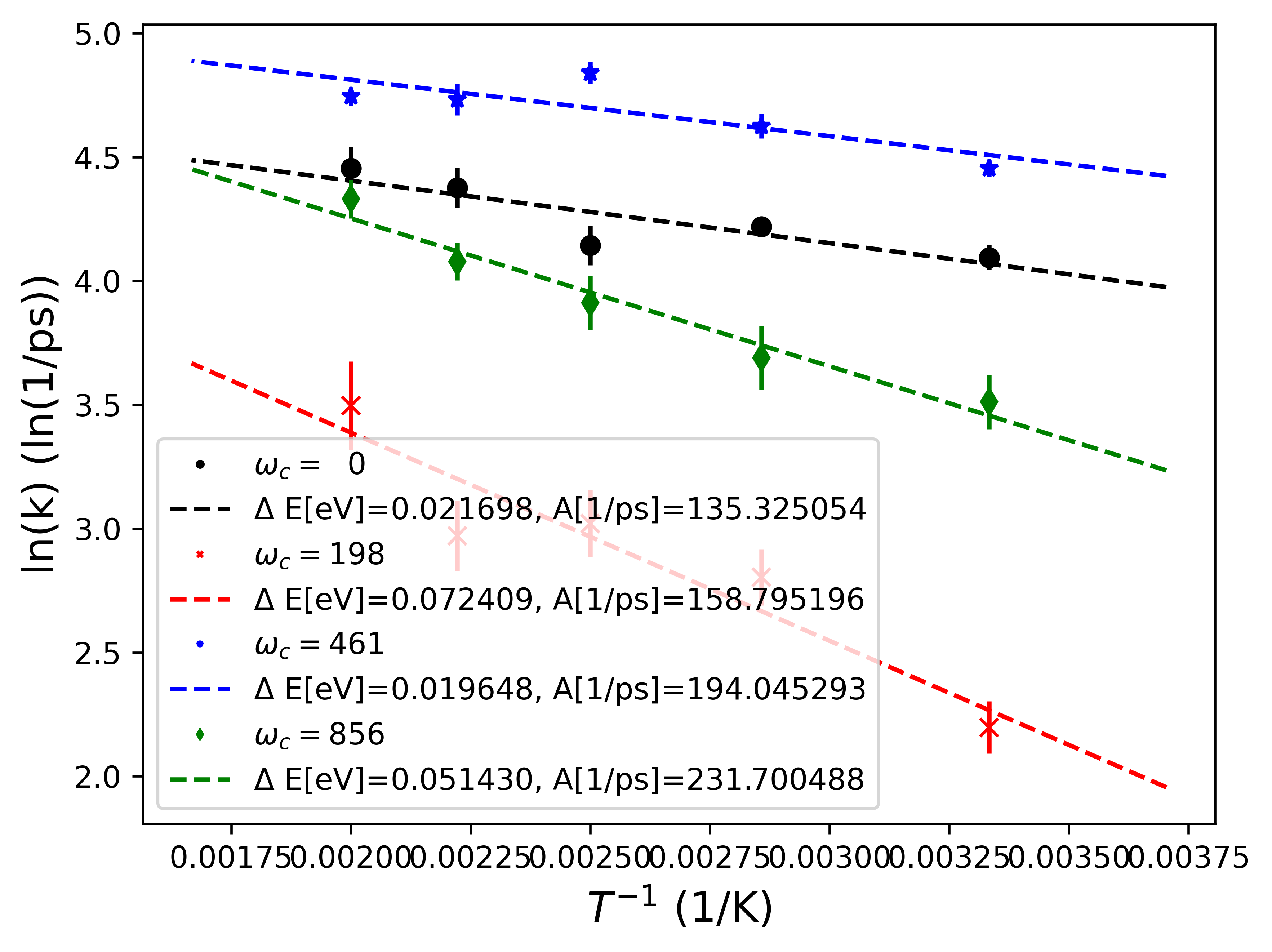}
    \caption{
    Arrhenius plot $k = A e^{-\Delta E / k_B T}$ for the unidirectional reaction $PTAF^- \rightarrow FtMeSi+PA^{-}$ obtained under NVE conditions.
    }
    \label{fig:Arrhenius}
\end{figure}

We point out that our initial state is energetically above the transition-state, i.e., the reaction is almost barrier-free. This results in the positive enthalpy shown in the Eyring plot (main document) and the low activation barrier in the Arrhenius plot \autoref{fig:Arrhenius}. We plan to provide a rigorous NVT equilibrium rate once a full GPU implementation is available.

\subsection{Consistency Checks}
\label{sec:natcomcheck}

\begin{figure}
    \centering
    \includegraphics[width=1.0\columnwidth]{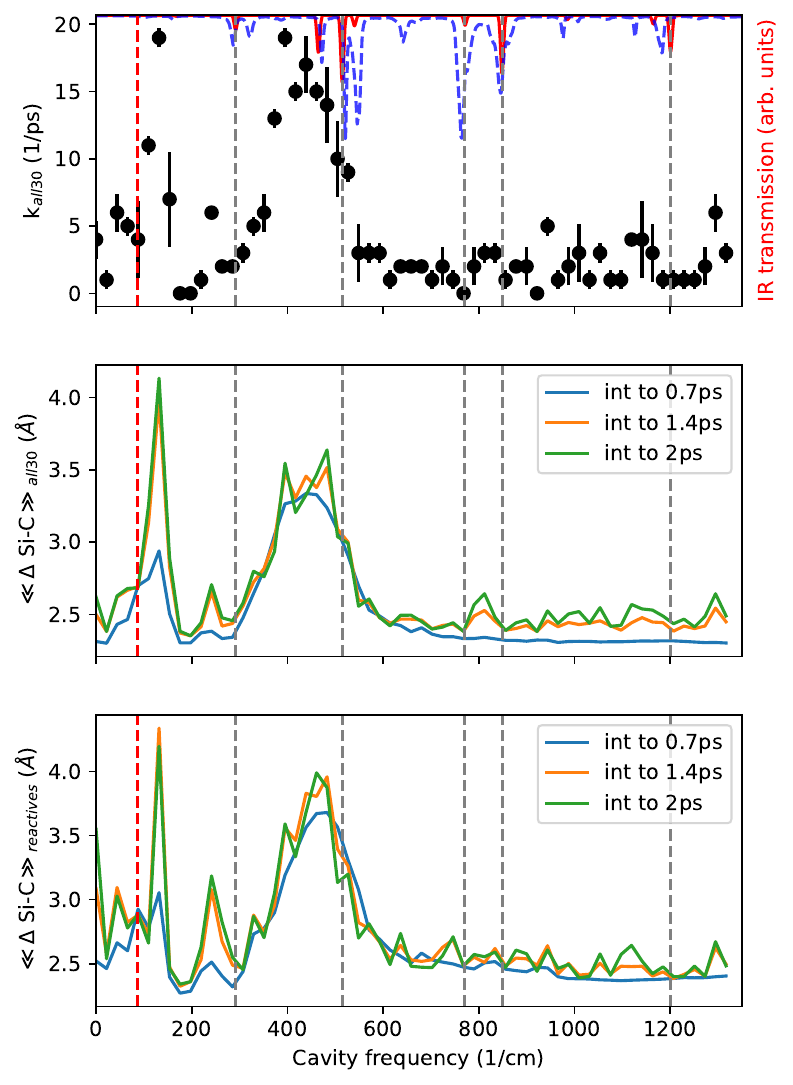}
    \caption{Top: Rate for the unidirectional reaction $PTAF^- \rightarrow FtMeSi+PA^{-}$ using the 30 initial configurations used in Ref.~\cite{schafer2021shining} and $g/\omega=1.132$, but propagated with our NEP based molecular dynamics calculator. Transmission spectrum obtained from Octopus at 0K, harmonic approximation (red solid), and using our NEP model and GPUMD at 400K NVE conditions (blue dashed). Vertical lines indicate characteristic features observed in Ref.~\cite{schafer2021shining}.\\
    Middle: Trajectory and time-averaged Si-C distance using all 30 trajectories for different time-intervals.\\
    Bottom: Trajectory and time-averaged Si-C distance for the subset of reactive trajectories for different time-intervals. This corresponds to the observable shown in Ref.~\cite{schafer2021shining}.  The relatively clear correlation with the rate supports the reliability of the averaged Si-C distance as indicator for rate changes. A sanity check for the difference between middle and bottom picture is, that the average Si--C distance at $\omega_c=0$ (outside cavity) is clearly higher for the subset of reactive trajectories which increases their average value. The overall differences are small, rate and Si--C distance show a strong correlation.}
    \label{fig:freqscanNatCom}
\end{figure}

Starting the same initial conditions as Ref.~\cite{schafer2021shining}, we have calculated rates and average Si--C distances. The obtained rates (Fig.~\ref{fig:freqscanNatCom}, top) are largely consistent with our observations in this work and are further discussed in the main text. Fig.~\ref{fig:freqscanNatCom} middle shows the Si--C distance averaged over the full ensemble and the specified time-domain. Fig.~\ref{fig:freqscanNatCom} bottom shows the same but only for the subset of trajectories that are reactive outside the cavity. The \textit{ab initio} calculations in Ref.~\cite{schafer2021shining} used a short integration domain of $0.7$~ps and utilized only the subset of trajectories reactive outside the cavity, i.e., the most consistent comparison is with the blue solid line in Fig.~\ref{fig:freqscanNatCom} bottom. Ref.~\cite{schafer2021shining} did not indicate any rate enhancing effect of the cavity, rate and Si--C distance are sufficiently correlated to draw a connection.

We re-optimized the TS and calculated the vibrational modes at the transition state. The frequency corresponding to the negative TS curvature is with $73.38$~cm$^{-1}$ close to previously reported values \cite{climent2020sn,li2020resonance,schafer2021shining}. 
In order to check if our NEP model informs the molecular dynamics simulations with the correct value, we \Cadd{fixed the Si-C atoms and} performed a short BFGS optimization with our NEP model starting from the TS structure obtained in ORCA. The calculated value of \Cadd{$69.5$}~cm$^{-1}$ \Cadd{(see SI Sec.~\ref{sec:vibfrequencies})} is in close agreement and suggests that any effect originating from the transition-state curvature should be correctly accounted for. Future work should investigate how the cavity Born-Oppenheimer approximation modifies our observation, as we would expect then a closer agreement with previous model and \textit{ab initio} calculations \cite{sun2022suppression,schafer2021shining}.

\subsection{Normal mode occupations}

\begin{figure*}[h]
    \centering
    \includegraphics[width=1.0\columnwidth]{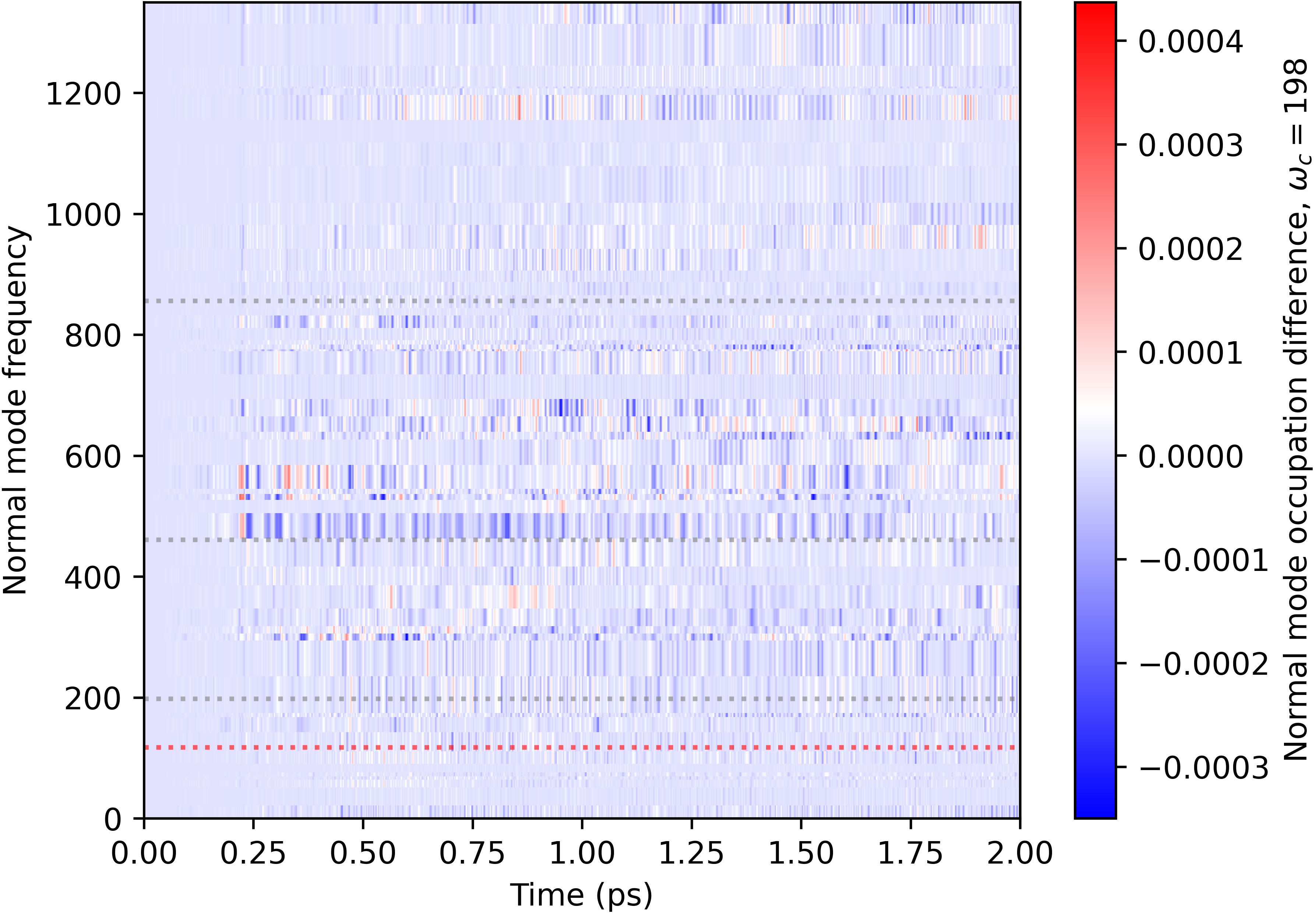} 
    \includegraphics[width=1.0\columnwidth]{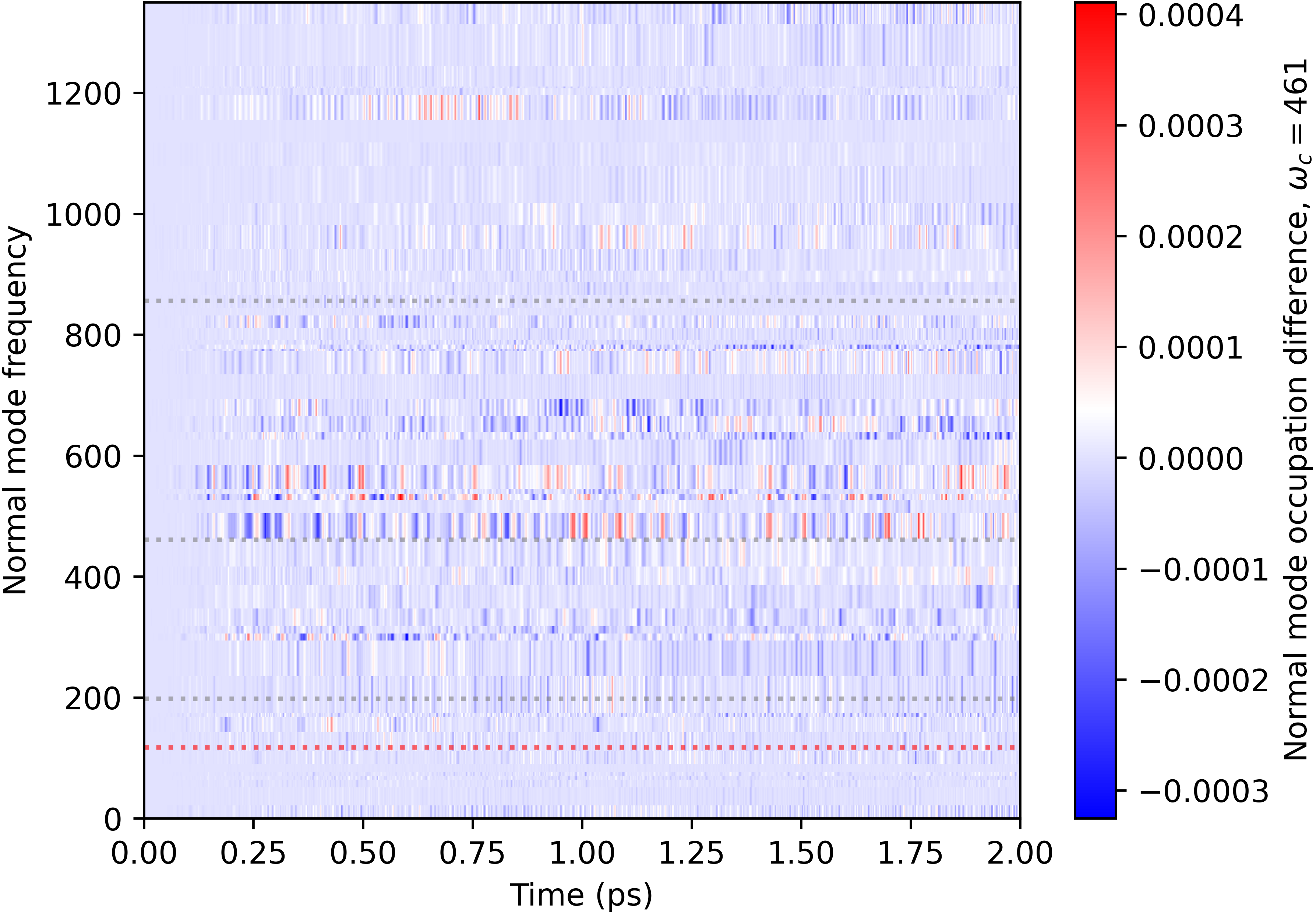}\\
    \includegraphics[width=1.0\columnwidth]{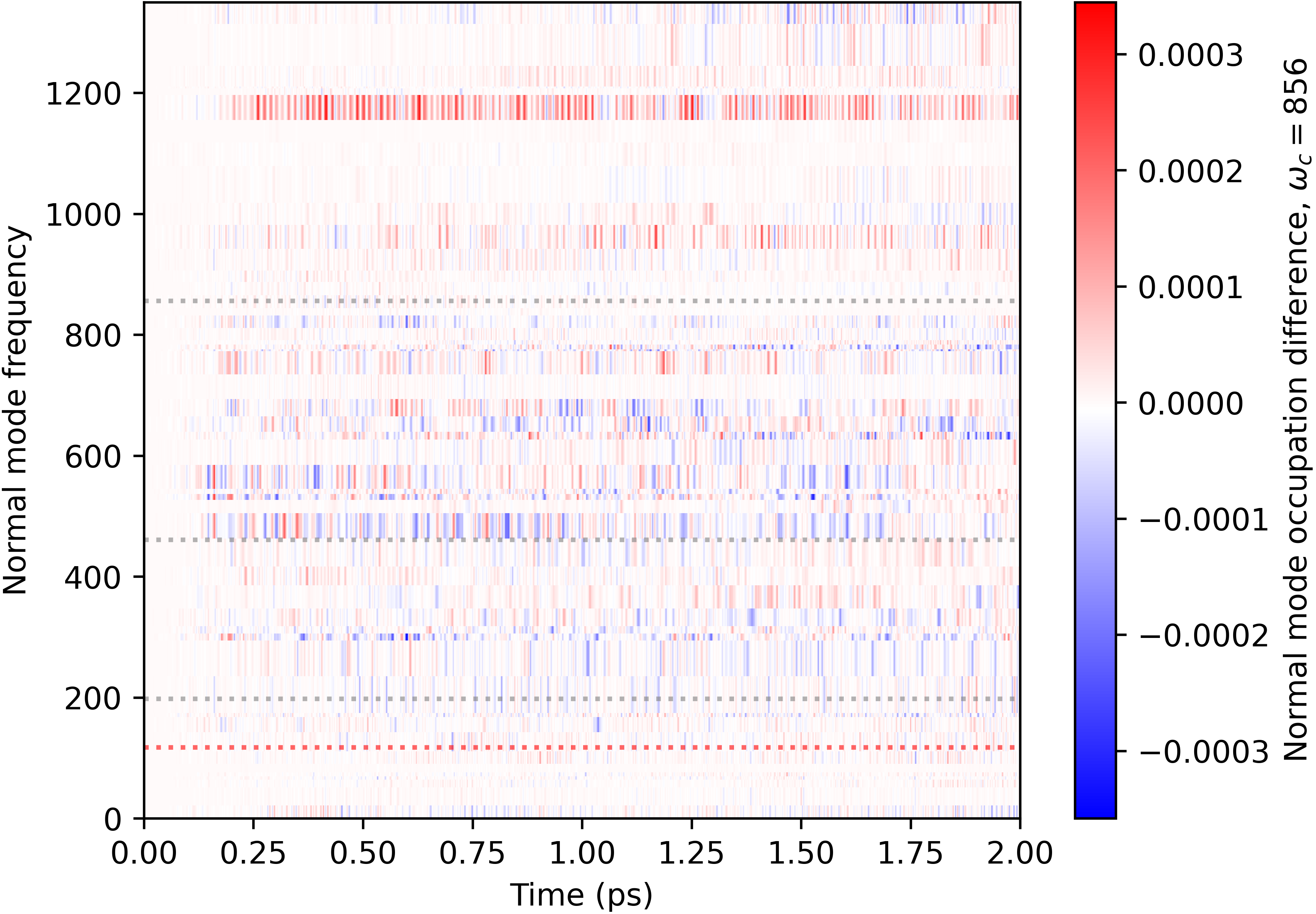}
    \includegraphics[width=1.0\columnwidth]{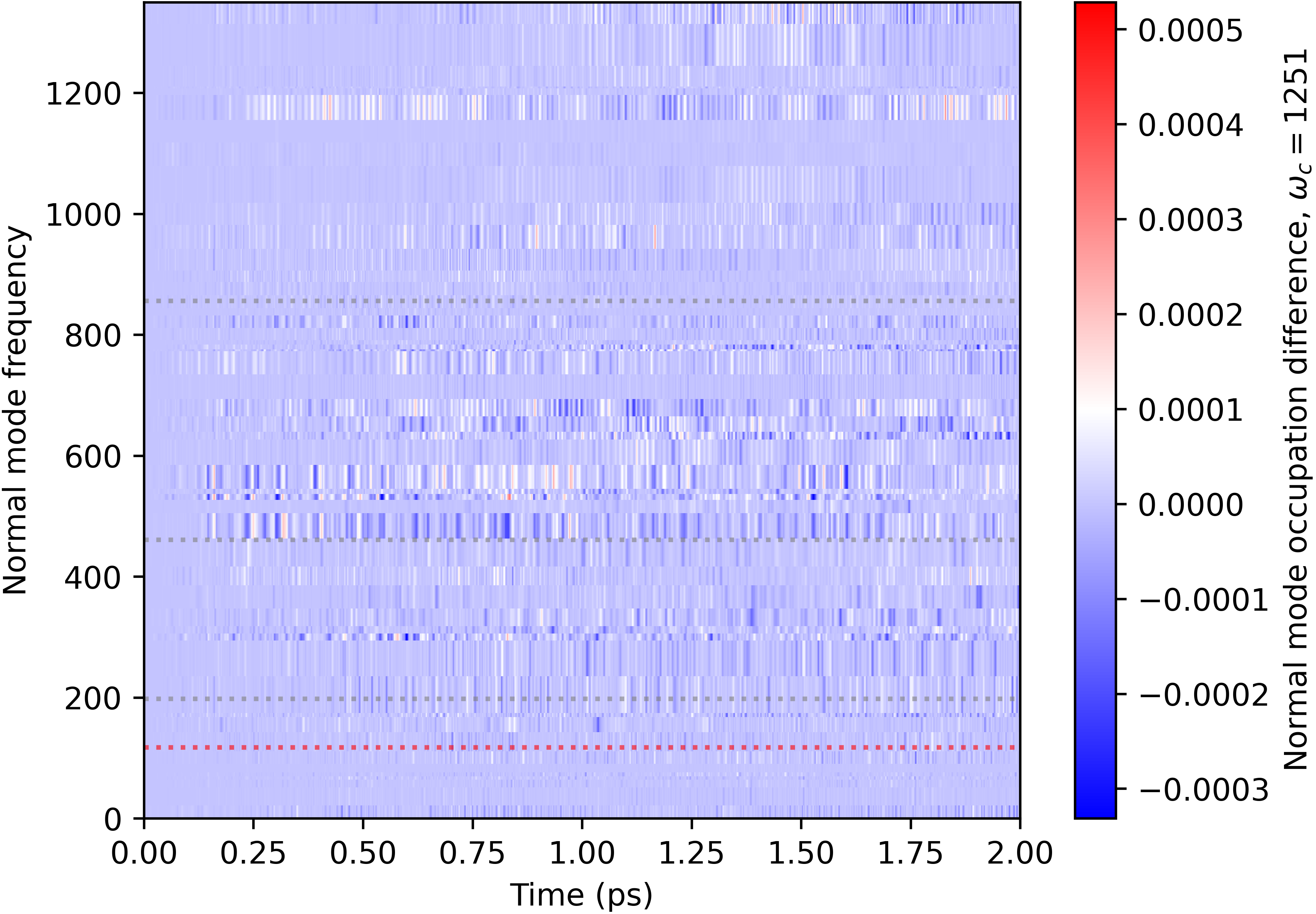}
    \caption{Top left: Difference in normal mode occupation for $\omega_c = 198$~cm$^{-1}$ and $g/\omega=1.132$ vs free space at 400 K.
    Top right: Difference in normal mode occupation for $\omega_c = 461$~cm$^{-1}$ and $g/\omega=1.132$ vs free space at 400 K.\\
    Bottom left: Difference in normal mode occupation for $\omega_c = 856$~cm$^{-1}$ and $g/\omega=1.132$ vs free space at 400 K.
    Bottom right: Difference in normal mode occupation for $\omega_c = 1251$~cm$^{-1}$ and $g/\omega=1.132$ vs free space at 400 K.}
    \label{fig:modediffs}
\end{figure*}

Fig.~\ref{fig:modediffs} illustrates the difference in normal mode occupation for 3 different choices of cavity frequency. The normal mode occupation $\textbf{o}$ is calculated by normalizing the force extracted from the trajectory at a given time $\textbf{f}(t) = \textbf{F}(t)/\lVert \textbf{F}(t) \lVert_2$ and projecting it onto the orthornormal set of normal mode forces $\textbf{o} = (\textbf{f}_{nm} \cdot \textbf{f}(t))^2,~ \textbf{f}_{nm}^T \cdot \textbf{f}_{nm} = \mathbb{1}$. 
Notice that the y-axis uses the frequency and we stretch the normal-mode occupation accordingly. The overall structure for $\omega_c = 856$~cm$^{-1}$ is comparable to Ref.~\cite{schafer2021shining}. For comparison, one should take into account, that the experimentally most relevant normal mode is located at 849~cm$^{-1}$ in Ref.~\cite{schafer2021shining} while it is located at 831~cm$^{-1}$ when using our NEP potential and the ASE internal vibrational mode calculator (0~K). Noticeable is the rather unaffected region around 750-1000~cm$^{-1}$ from which only the optically active modes at 770 and 831~cm$^{-1}$ stand out. The domain between 450-550~cm$^{-1}$ is quite pronounced (similar to Ref.~\cite{schafer2021shining}).

\subsection{\Cadd{Si--C stretching contribution in vibrational modes}}

\begin{figure}[h]
    \centering
    \includegraphics[width=1.0\columnwidth]{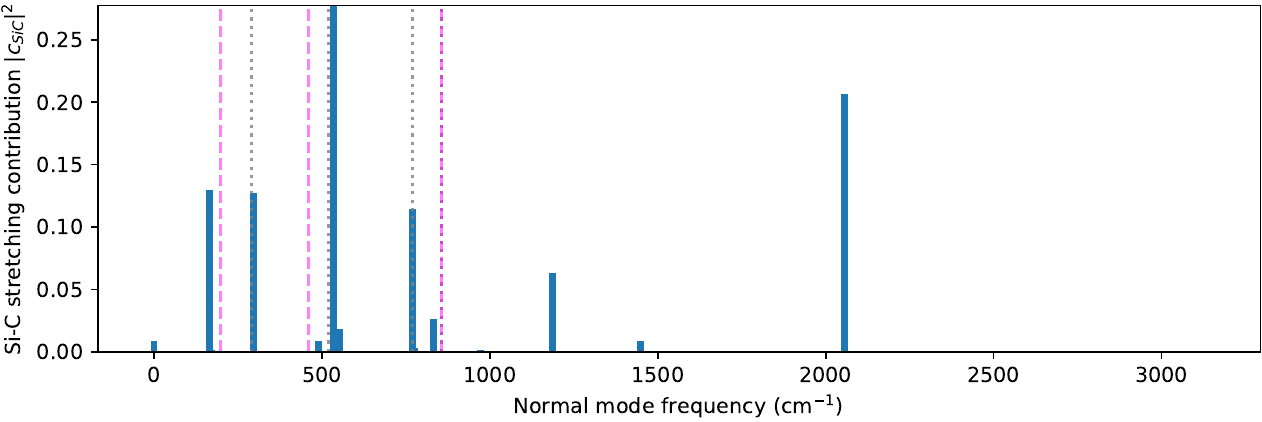} 
    \caption{Si--C contribution to vibrational normal modes. Obtained by projecting a force $\textbf{f}_{SiC}=-\frac{1}{\sqrt{2}}\textbf{e}^{Si}_{x}+\frac{1}{\sqrt{2}}\textbf{e}^{C}_{x}$ on the normal-mode forces (equivalent to \autoref{fig:modediffs}).}
    \label{fig:siccontribution}
\end{figure}

\Cadd{
Since the reactive step involves breaking the Si--C bond, the contribution of Si--C stretchings in the normal modes is an important indicator for the expected impact on the reaction when energy is redistributed between optically active modes. \autoref{fig:siccontribution} demonstrates that, in agreement with Fig.~2C from the main manuscript, the Si--C stretching contributions are located foremost in the energy-window between 160~cm$^{-1}$ and 840~cm$^{-1}$, with the exception of the C=C bond stretching around 1200~cm$^{-1}$ and a high-energy mode beyond above 2000~cm$^{-1}$.
Nonetheless, this supports our argumentation that the reactive modes that could be effected of dynamic electronic polarization are localized below 840~cm$^{-1}$ which explains why changes at higher frequencies disappear.
}

\subsection{Unit conversion between atomic units and ASE units}

The dimensionless ratio $g_0/\omega_c$ determines our coupling strength. It should be noted that using the implemented cavity-calculator requires the coupling $\lambda = \frac{1}{\sqrt{\varepsilon_0 V_c}}$ in ASE units. Our scripts handle this conversion automatically but we provide a brief discussion of the relevant conversion to facilitate reproduction from independent researchers.
Since $g_0/\omega_c = [\mu] \sqrt{1/\hbar\omega_c2\varepsilon_0V_c}$, where $[\mu]$ denotes the units of the dipole moment, and thus $[g_0/\omega_c] = [1] = [\mu] [\lambda] / \sqrt{[\hbar \omega_c]}$ as well as  $[\lambda] = \sqrt{[\hbar \omega_c]}/[\mu]$.
Furthermore, $\lambda^{au} = \frac{g_0^{au}}{\omega_c^{au}} \sqrt{2\omega_c^{au}}$ and we finally arrive at 
$\lambda^{ase} =  \frac{g_0^{au}}{\omega_c^{au}} \sqrt{2\omega_c^{au}} \frac{\sqrt{[energy]_{au~to~ase}}}{[length]_{au~to~ase}}$ ($[x]_{au~to~ase}$ are conversions between units) which can be directly related to Ref.~\cite{schafer2021shining} by taking $ \frac{g_0^{au}}{\omega_c^{au}} = 1.132$.

\subsection{\Cadd{NEP model using a smaller electronic basis}}
\label{sec:631g}

\begin{figure}
    \centering
    \includegraphics{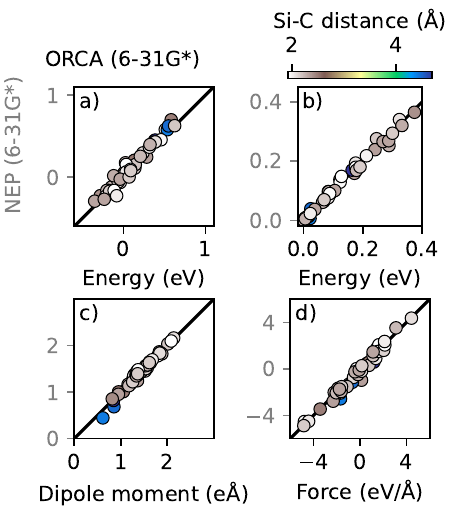}
    \caption{Parity plots for the alternative NEP model trained on the 6-31G* basis set versus ab initio calculations performed with ORCA and the 6-31G* basis set.
    Comparison of (a) potential energies of the electronic system (we have subtracted \qty{-22206.5}{\electronvolt} from the values), (b) kinetic energies of the electronic system, (c) dipole moments and (d) the (electronic + cavity) force acting on the Si-C bond projected on the bond vector.}
    \label{fig:parity_plots_NEP_631_ORCA}
\end{figure}

\Cadd{
As detailed in Sec.~\ref{sec:train}, we started training a second NEP model based on DFT calculations (using ORCA) with the smaller 6-31G* basis set.
Dipole moments and stretched configurations are less reliable when using the small 6-31G* basis, such that the following presentation should be consider with caution. As shown in \autoref{fig:parity_plots_NEP_631_ORCA}, the trained NEP models based on the 6-31G* calculations accurately reproduce the forces obtained from ORCA.
}

\Cadd{
The smaller basis results in a lower reaction barrier and thus a quick saturation of the limited set of trajectories.
Nonetheless, we can use those NEP models to investigate if the dynamic electronic polarization remains an important components resulting in comparable deviations.
}

\begin{figure}
    \centering
    \includegraphics[width=1.0\columnwidth]{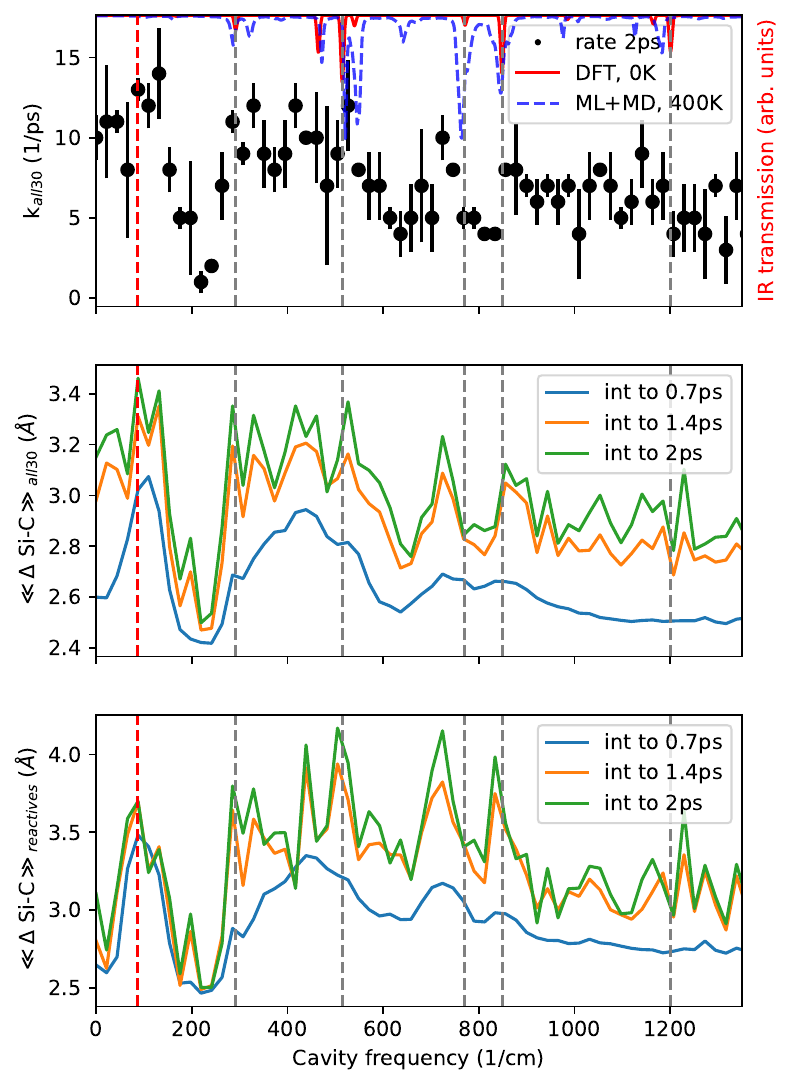}
    \caption{Rate and Si--C averages corresponding to \autoref{fig:freqscanNatCom} but obtained using the 6-31G* trained NEP models.
    Top: Rate for the unidirectional reaction $PTAF^- \rightarrow FtMeSi+PA^{-}$ using the 30 initial configurations used in Ref.~\cite{schafer2021shining} and $g/\omega=1.132$, but propagated with our NEP based molecular dynamics calculator. Transmission spectrum obtained from Octopus at 0K, harmonic approximation (red solid), and using our NEP model and GPUMD at 400K NVE conditions (blue dashed). Vertical lines indicate characteristic features observed in Ref.~\cite{schafer2021shining}.\\
    Middle: Trajectory and time-averaged Si-C distance using all 30 trajectories for different time-intervals.\\
    Bottom: Trajectory and time-averaged Si-C distance for the subset of trajectories used in \autoref{fig:freqscanNatCom}.}
    \label{fig:freqscanNatCom_631}
\end{figure}

\Cadd{
\autoref{fig:freqscanNatCom_631} presents reaction rate constants and Si--C distances obtained from starting ML+MD calculation using the 6-31G* trained NEP models.
Two aspects require additional discussion:
\begin{enumerate}
    \item The overall shape is consistent with \autoref{fig:freqscanNatCom} for the average change in Si--C distance in the first 700~fs (blue lines in middle and bottom plot). The catalysing character around 450/cm is smaller and seems to be broader, such that the domain around 800/cm obtains additional weight, better emphasizing the resonant dependence. The overall effect remains small at larger frequencies.
    \item All frequencies besides 200/cm quickly reach comparable number of products as obtained at the catalysing frequencies explained in the first aspect, i.e., the lower barrier allows the non-catalyzed trajectories to catch up. The only feature that is clearly standing out after 2~ps is the strong inhibition at 200/cm (in Si--C averages and rate). 
\end{enumerate}
The basic qualitative behavior of both ML+MD investigations is consistent in the first 700~fs.
However, the quick increase at all frequencies (besides 200/cm) implies that only few features remain to play a role after longer propagation time or broader sampling, i.e., only the features reported in Fig.~2C of the paper can be expected to remain relevant after proper sampling.
Nonetheless, that the qualitative trend of both ML+MD investigations (up to 700~fs) is consistent and qualitatively contradicts the QEDFT calculations (in the same 700~fs domain) supports our conclusions in the main paper. 
}

\subsection{\Cadd{Vibrational frequencies}}
\label{sec:vibfrequencies}

\begin{figure*}[h]
    \centering
    \includegraphics[width=1.0\textwidth]{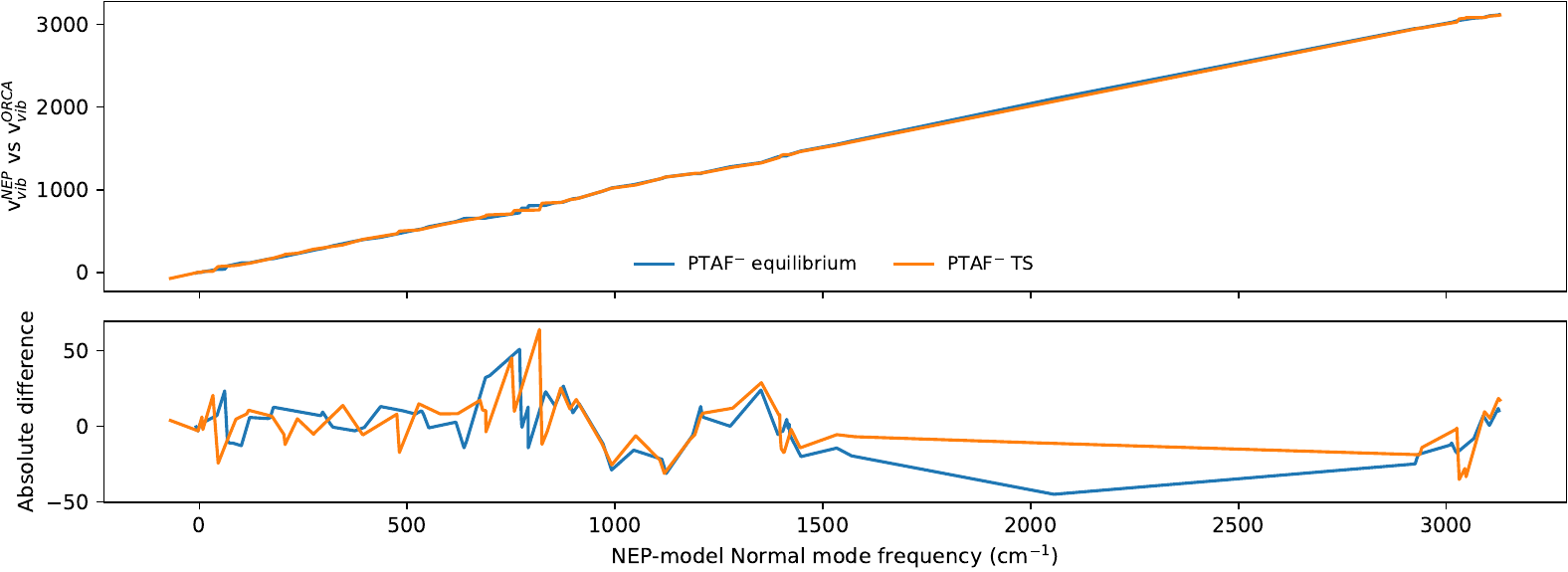}
    \caption{Normal-mode frequencies and their difference obtained at the intermediate state (blue) and the transition-state (orange) when using our NEP model or ORCA (both based on def2-TZVP basis).}
    \label{fig:vibfrequencies}
\end{figure*}

\Cadd{
\autoref{fig:vibfrequencies} presents the normal-mode frequencies and their difference obtained at the intermediate state (blue) and the transition-state (orange) when using our NEP model or ORCA (both based on def2-TZVP basis). The modes are sorted by their frequency.
}

\Cadd{
The ORCA transition-state (TS) geometry was obtained by reoptimizing an initial gues using the "OptTS" flag. The hessian was calculated at the beginning of the optimization to ensure reliable convergence.
Using our NEP model, the TS and its frequencies have been calculated by using the final TS structures from ORCA, fixing the positions of the two relevant Si-C atoms, performing a BFGS optimization to reduce forces down to a maximum value of $10^{-12}$, and calculating the vibrational modes of this relaxed TS structure.
The lowest TS frequency of NEP model and ORCA calculation are with $69.5$ and $73.38$~cm$^{-1}$ in close agreement.
}

\Cadd{
In addition, we provide below the explicit normal-mode frequencies.
}

\paragraph{Intermediate state vibrational frequencies NEP model:}
\begin{verbatim}
---------------------
  #    meV     cm^-1
---------------------
  0    1.0i      7.8i
  1    0.4i      3.1i
  2    0.2i      1.9i
  3    0.1i      0.5i
  4    0.0i      0.1i
  5    0.0i      0.0i
  6    0.2       1.4
  7    5.3      42.6
  8    7.6      61.0
  9    8.4      67.5
 10    9.0      72.9
 11    9.9      79.9
 12   12.5     101.1
 13   15.0     121.0
 14   20.5     165.1
 15   21.1     170.3
 16   22.1     178.4
 17   36.2     292.4
 18   36.8     296.8
 19   39.1     315.3
 20   39.8     321.4
 21   46.4     374.1
 22   49.2     397.1
 23   54.1     436.2
 24   60.8     490.1
 25   64.4     519.7
 26   66.3     534.7
 27   66.9     539.3
 28   68.4     552.0
 29   76.5     617.3
 30   79.0     636.8
 31   79.7     642.5
 32   85.4     688.6
 33   86.6     698.7
 34   95.5     770.5
 35   96.1     775.1
 36   96.3     776.5
 37   98.1     791.2
 38   98.1     791.4
 39  103.0     831.1
 40  103.3     833.4
 41  106.1     855.6
 42  108.6     875.9
 43  111.4     898.8
 44  113.3     913.8
 45  120.4     971.3
 46  123.0     991.8
 47  129.6    1045.1
 48  138.0    1112.9
 49  139.3    1123.2
 50  147.2    1187.1
 51  149.6    1206.3
 52  149.9    1209.2
 53  150.3    1212.1
 54  158.4    1277.3
 55  167.5    1351.3
 56  172.6    1392.3
 57  172.8    1393.7
 58  174.1    1404.0
 59  175.2    1413.3
 60  175.6    1416.6
 61  176.0    1419.4
 62  176.1    1420.6
 63  179.5    1447.5
 64  190.2    1533.7
 65  194.6    1569.2
 66  254.9    2056.3
 67  362.7    2925.3
 68  363.2    2929.3
 69  363.7    2933.2
 70  373.2    3009.7
 71  373.6    3013.1
 72  373.6    3013.3
 73  374.8    3023.2
 74  375.0    3024.2
 75  375.1    3025.2
 76  380.2    3066.6
 77  383.0    3089.1
 78  384.8    3103.9
 79  387.5    3125.2
 80  387.7    3127.1
\end{verbatim}
\paragraph{Intermediate state vibrational frequencies ORCA:}
\begin{verbatim}
-----------------------
VIBRATIONAL FREQUENCIES
-----------------------
   0:         0.00 cm**-1
   1:         0.00 cm**-1
   2:         0.00 cm**-1
   3:         0.00 cm**-1
   4:         0.00 cm**-1
   5:         0.00 cm**-1
   6:        -7.42 cm**-1 ***imaginary mode***
   7:        35.37 cm**-1
   8:        37.55 cm**-1
   9:        77.07 cm**-1
  10:        84.03 cm**-1
  11:        90.97 cm**-1
  12:       113.94 cm**-1
  13:       115.14 cm**-1
  14:       159.98 cm**-1
  15:       164.68 cm**-1
  16:       165.73 cm**-1
  17:       285.36 cm**-1
  18:       287.37 cm**-1
  19:       313.75 cm**-1
  20:       321.97 cm**-1
  21:       377.11 cm**-1
  22:       397.83 cm**-1
  23:       423.11 cm**-1
  24:       479.85 cm**-1
  25:       511.57 cm**-1
  26:       524.41 cm**-1
  27:       531.28 cm**-1
  28:       553.05 cm**-1
  29:       614.53 cm**-1
  30:       651.06 cm**-1
  31:       652.71 cm**-1
  32:       656.27 cm**-1
  33:       665.17 cm**-1
  34:       719.49 cm**-1
  35:       773.21 cm**-1
  36:       777.17 cm**-1
  37:       778.38 cm**-1
  38:       805.68 cm**-1
  39:       809.53 cm**-1
  40:       810.52 cm**-1
  41:       842.00 cm**-1
  42:       849.28 cm**-1
  43:       889.89 cm**-1
  44:       898.81 cm**-1
  45:       982.89 cm**-1
  46:      1020.73 cm**-1
  47:      1060.90 cm**-1
  48:      1134.72 cm**-1
  49:      1154.47 cm**-1
  50:      1192.98 cm**-1
  51:      1193.18 cm**-1
  52:      1200.73 cm**-1
  53:      1206.08 cm**-1
  54:      1277.25 cm**-1
  55:      1327.29 cm**-1
  56:      1397.82 cm**-1
  57:      1398.03 cm**-1
  58:      1407.21 cm**-1
  59:      1408.67 cm**-1
  60:      1417.89 cm**-1
  61:      1417.98 cm**-1
  62:      1426.23 cm**-1
  63:      1467.57 cm**-1
  64:      1548.10 cm**-1
  65:      1588.76 cm**-1
  66:      2101.33 cm**-1
  67:      2950.21 cm**-1
  68:      2950.59 cm**-1
  69:      2951.78 cm**-1
  70:      3022.11 cm**-1
  71:      3024.15 cm**-1
  72:      3024.48 cm**-1
  73:      3040.26 cm**-1
  74:      3040.60 cm**-1
  75:      3042.77 cm**-1
  76:      3074.66 cm**-1
  77:      3082.10 cm**-1
  78:      3103.34 cm**-1
  79:      3113.39 cm**-1
  80:      3116.81 cm**-1
\end{verbatim}

\paragraph{Transition state vibrational frequencies NEP model:}
\begin{verbatim}
---------------------
  #    meV     cm^-1
---------------------
  0    8.6i     69.5i
  1    0.4i      3.0i
  2    0.0i      0.1i
  3    0.0i      0.1i
  4    0.1       0.5
  5    0.7       5.6
  6    0.8       6.2
  7    1.1       8.5
  8    4.0      32.7
  9    4.2      33.9
 10    5.6      45.3
 11    7.6      61.5
 12   10.9      88.1
 13   14.2     114.4
 14   14.7     118.5
 15   20.8     167.9
 16   21.7     175.1
 17   25.2     203.6
 18   25.6     206.3
 19   29.2     235.9
 20   34.1     275.2
 21   34.3     276.9
 22   42.8     345.4
 23   48.8     393.3
 24   58.9     475.2
 25   59.7     481.3
 26   65.5     528.2
 27   71.8     579.3
 28   77.1     621.7
 29   83.8     675.5
 30   84.5     681.7
 31   85.4     689.2
 32   85.5     689.4
 33   86.1     694.6
 34   93.2     751.6
 35   94.0     757.9
 36   95.2     768.0
 37  101.5     818.4
 38  101.8     821.4
 39  102.2     824.5
 40  103.8     837.4
 41  104.8     845.3
 42  107.8     869.6
 43  110.5     891.6
 44  112.4     907.0
 45  120.1     968.5
 46  123.0     992.3
 47  130.2    1049.9
 48  137.4    1108.2
 49  138.7    1118.8
 50  146.2    1179.5
 51  147.8    1192.4
 52  149.4    1205.3
 53  150.1    1211.0
 54  159.1    1283.3
 55  167.7    1352.3
 56  173.0    1395.0
 57  173.1    1396.5
 58  173.2    1397.2
 59  173.6    1400.0
 60  174.3    1406.2
 61  174.5    1407.6
 62  176.5    1423.8
 63  179.2    1445.5
 64  190.2    1533.9
 65  195.5    1577.1
 66  248.6    2005.5
 67  364.0    2936.1
 68  364.6    2940.6
 69  364.7    2941.6
 70  374.8    3022.6
 71  375.0    3024.9
 72  375.1    3025.6
 73  375.8    3031.2
 74  377.5    3045.0
 75  377.9    3047.8
 76  380.4    3068.3
 77  383.4    3092.3
 78  385.0    3105.2
 79  387.5    3125.7
 80  388.0    3129.3
\end{verbatim}

\paragraph{Transition state vibrational frequencies ORCA:}
\begin{verbatim}
-----------------------
VIBRATIONAL FREQUENCIES
-----------------------
   0:         0.00 cm**-1
   1:         0.00 cm**-1
   2:         0.00 cm**-1
   3:         0.00 cm**-1
   4:         0.00 cm**-1
   5:         0.00 cm**-1
   6:       -73.38 cm**-1 ***imaginary mode***
   7:        10.65 cm**-1
   8:        12.17 cm**-1
   9:        17.68 cm**-1
  10:        69.77 cm**-1
  11:        74.63 cm**-1
  12:        83.36 cm**-1
  13:       106.14 cm**-1
  14:       107.85 cm**-1
  15:       160.73 cm**-1
  16:       168.58 cm**-1
  17:       209.02 cm**-1
  18:       218.38 cm**-1
  19:       230.93 cm**-1
  20:       280.51 cm**-1
  21:       281.42 cm**-1
  22:       331.60 cm**-1
  23:       399.03 cm**-1
  24:       467.18 cm**-1
  25:       498.66 cm**-1
  26:       513.31 cm**-1
  27:       571.07 cm**-1
  28:       613.36 cm**-1
  29:       658.26 cm**-1
  30:       671.00 cm**-1
  31:       678.83 cm**-1
  32:       693.12 cm**-1
  33:       694.55 cm**-1
  34:       706.36 cm**-1
  35:       748.03 cm**-1
  36:       749.11 cm**-1
  37:       754.29 cm**-1
  38:       799.53 cm**-1
  39:       836.44 cm**-1
  40:       840.56 cm**-1
  41:       841.53 cm**-1
  42:       844.28 cm**-1
  43:       880.07 cm**-1
  44:       889.23 cm**-1
  45:       979.89 cm**-1
  46:      1017.88 cm**-1
  47:      1056.16 cm**-1
  48:      1130.63 cm**-1
  49:      1149.99 cm**-1
  50:      1188.63 cm**-1
  51:      1197.92 cm**-1
  52:      1199.54 cm**-1
  53:      1202.36 cm**-1
  54:      1271.28 cm**-1
  55:      1323.42 cm**-1
  56:      1387.79 cm**-1
  57:      1388.71 cm**-1
  58:      1401.19 cm**-1
  59:      1414.79 cm**-1
  60:      1423.58 cm**-1
  61:      1424.43 cm**-1
  62:      1425.78 cm**-1
  63:      1459.76 cm**-1
  64:      1539.43 cm**-1
  65:      1584.00 cm**-1
  66:      2016.38 cm**-1
  67:      2954.93 cm**-1
  68:      2955.41 cm**-1
  69:      2955.64 cm**-1
  70:      3024.72 cm**-1
  71:      3026.47 cm**-1
  72:      3026.73 cm**-1
  73:      3066.43 cm**-1
  74:      3072.94 cm**-1
  75:      3081.19 cm**-1
  76:      3081.98 cm**-1
  77:      3082.63 cm**-1
  78:      3099.89 cm**-1
  79:      3107.26 cm**-1
  80:      3111.92 cm**-1
\end{verbatim}


\bibliography{references}